\documentclass{JFM-FLM_Au}

\usepackage{pgfplots}
\pgfplotsset{compat=1.9}

\usepackage{import}     % needed for \input{} with relative paths in .pdf_tex
\usepackage{subcaption}     %% To facilitate subfigures
\usepackage{float}          %% for positioning the tables
\usepackage[section]{placeins}  % keeps floats within sections

\usetikzlibrary{calc}

\definecolor{mygreen}{RGB}{119,172,48}

\usepackage{upgreek}

\lefttitle{S. Vayala, K. Ramachandra, K. Abhishek, N. R. Vadlamani and R. Sriram}
\righttitle{Unsteadiness in bow shock-induced separation}

\title{On the governing mechanism of unsteadiness in bow shock-induced three-dimensional separation}
\author{Siva Vayala\aff{1}, K. Ramachandra\aff{2}, K. Abhishek\aff{1}, Nagabhushana Rao Vadlamani\aff{1} \and R. Sriram\aff{1$\dag$}}

\affiliation{\aff{1}Department of Aerospace Engineering, Indian Institute of Technology Madras, Chennai, India
\aff{2}Department of Engineering, University of Cambridge, Cambridge, United Kingdom}

\corresau{$\dag$ R. Sriram, \email{r.sriram@iitm.ac.in}}

\begin{document}
\maketitle

% \makeatletter
% \def\ps@titlepage{%
%   \def\@oddhead{\normalfont\fontsize{9}{11}\selectfont{\itshape This draft was prepared using the LaTeX style file belonging to the Journal of Fluid Mechanics}}%
%   \def\@evenhead{}%
%   \def\@oddfoot{}%
%   \def\@evenfoot{}%
% }
% \makeatother

% \makeatletter
% \let\ps@plain\ps@fancy
% \makeatother

\begin{abstract}

We investigate the driving mechanism of low-frequency unsteadiness in bow shock-turbulent boundary layer interactions due to protuberances. Wind tunnel experiments are conducted at a freestream Mach number of 2.87 with protuberances of different shapes and sizes. From time-resolved surface pressure measurements and schlieren imaging, the unsteadiness is characterized by low-frequency shock oscillations, with a Strouhal number of $St_{\delta}\sim 0.01$ based on the boundary layer thickness ($\delta$), while the separated region exhibits predominantly mid-frequency pressure oscillations, with $St_{\delta} \sim 0.1$. Mid-span separation length, $L_{sep}$, is identified as a key parameter in determining time and length scales of shock oscillations. Further details of the interaction are examined through compressible adaptive detached eddy simulations for one particular case, viz.,the cubical protuberance of side 15 mm. A detailed modal analysis using proper orthogonal decomposition (POD) is performed with the 3-D data from computations. Flapping of shock-foot about mid-span was apparent, over and above the coherent to-and-fro oscillations, with the dominance of anti-symmetric mode in the POD of wall pressure fluctuations. The motion of the shock foot is initiated near mid-span, while the shock foot at other spanwise locations lags behind. The flap and asymmetries are related to the spanwise extent of reverse flow. From the reconstructed 3-D flow field using low-frequency modes, along with corroborating observations from the two-point correlations, it is inferred that the imbalance and time lag between the mass injected into the separated region at reattachment and the mass leaving spanwise at the horseshoe vortex core govern the observed shock motion.

% It establishes the centerline separation length, ������ ��, as a key parameter558
% in determining shock oscillation frequencies and amplitudes. 
\end{abstract}

\begin{keywords} High-speed flow, Compressible boundary layers, Boundary layer separation

% Authors should not enter keywords on the manuscript, as these must be chosen by the author during the online submission process and will then be added during the typesetting process (see \href{https://www.cambridge.org/core/journals/journal-of-fluid-mechanics/information/list-of-keywords}{Keyword PDF} for the full list).  Other classifications will be added at the same time. \\ \\
\end{keywords}

%{\bf MSC Codes }  {\it(Optional)} Please enter your MSC Codes here

\section{Introduction}
\label{sec1:intro}
Shock wave boundary layer interactions (SBLI) could result in flow separation and unsteadiness, imposing unsteady pressure and heating loads, which in turn affect the performance of high-speed systems. At the turn of the century, \cite{dolling2001fifty} identified unsteadiness as one of the two important unresolved issues in SBLI. The unsteadiness in discussion here is the often observed large-scale to-and-fro excursion of separation/reflected shock about the mean position, exhibiting low characteristic frequencies at least two orders of magnitude lower than the characteristic incoming turbulent boundary layer frequencies and one order of magnitude lower than the characteristic flapping frequencies in the separated shear layer. Such low-frequency shock oscillations can lead to structural catastrophe in high-speed flow applications by exposing vehicle surfaces to fatigue aero-thermal loads at frequencies close to their natural frequencies. The consequences can also include intake buzz, unstart, etc., issues in supersonic air-breathing systems, thereby affecting their operational efficiency. This situation has motivated the decades-long debate on the source of the low-frequency unsteadiness; some works have argued in favor of `upstream mechanism’, i.e., the incoming turbulent boundary layer being the source of unsteadiness, while others have attributed unsteadiness to ‘downstream mechanism’, i.e., from within the interaction region itself. The experimental reports by \cite{andreopoulos1987some} and by \cite{thomas1994mechanism} were among the earliest, speculating the role of upstream and downstream sources in driving the low-frequency unsteadiness, respectively.

With advancements in flow diagnostics and computations, the aspect of unsteadiness in SBLI has received a significant resolution over the last two decades, at least in the case of nominally two-dimensional canonical configurations. \cite{andreopoulos1987some} observed that large-scale shock motion occurs at velocities and frequencies similar to those of velocity fluctuations and turbulent bursts in the upstream boundary layer, thereby speculating that it may be the source of shock motion. \cite{beresh2002relationship}, through their studies on compression ramps at Mach 5, suggested that the shock motions are strongly correlated with the velocity fluctuations in the upstream boundary layer. They showed that positive velocity fluctuations, having higher momentum, correlate with the downward movement of the separation shock, while negative fluctuations are correlated with the upstream motion of the separation shock. By means of wide field PIV and planar laser scattering measurements in streamwise-spanwise planes, \cite{ganapathisubramani2007effects, ganapathisubramani2009low} identified the presence of long coherent structures (superstructures) in the incoming boundary layer, 40 times the boundary layer thickness ($\delta$), which affected the separated flow induced by a compression corner. \cite{humble2009three} too investigated the role of such incoming superstructures for the case of impinging shock interaction, using volumetric measurements rather than point/planar measurements, using tomographic PIV. Regarding superstructures, \cite{touber2009large} made interesting observations through large eddy simulations (LES) of oblique shock reflection case; even with relatively small turbulent streaks in the incoming boundary layer, the interaction exhibited low-frequency shock oscillations, suggesting that the origin of low-frequency unsteadiness is not necessarily from the coherent structures, although the structures can have some influence as a forcing. Recently, \cite{porter2019selective} suggested that large-scale pulsations of the separation region are related to the selective influence of large-scale turbulent structures near the wall. Near-wall turbulent structures in the boundary layer act as a forcing mechanism, resulting in a global mode of separation bubble. A noteworthy feature in most works emphasizing the role of incoming turbulence on unsteadiness is the relatively small separation bubble, close to incipient separation.

Many other works argue in favour of downstream mechanism. Through time-resolved pressure measurements in the interaction zone (for impinging shock interaction), and from interpreting the results from other reported experiments in the literature on interaction unsteadiness in different configurations, \cite{dussauge2006unsteadiness} suggested the source of unsteadiness to be from within the interaction itself, speculating that the three-dimensional structure (viewed using streamwise-spanwise planar PIV) of the separated region may be a source of the unsteadiness; the shedding of vortices downstream of the shock could be the reason. In the spectra of unsteady surface pressures at the shock foot location, the peak energy was at Strouhal number, defined based on interaction length (L), $St_{L} \sim 0.02-0.06$. Within the recirculation region, the peak energy was at $St_{L} \sim 0.5$, which is also typically the orders of magnitude of the mixing layer oscillations (flapping). Further, the high coherence between these signals at the low frequency range (of shock oscillations) and low correlations of these signals with pressure fluctuations upstream of interaction zone, strongly pointed towards the `downstream mechanism’. Such spectra and coherences in surface pressure are typically reported over wide range of Mach numbers for any canonical 2-D SBLI with separation, such as with impinging shock \citep{erengil1993effects, dussauge2006unsteadiness, dupont2006space, ringuette2009experimental} or compression corner \citep{erengil1991unsteady, thomas1994mechanism, pasquariello2017unsteady, manisankar2023control}. 

Direct numerical simulation (DNS) studies have also pointed towards the downstream causes for the (separation) shock unsteadiness. \cite{pirozzoli2006direct} analyzed by means of DNS the case of impinging shock interaction at Mach 2.25, with mild separation. It was proposed that an acoustic resonance mechanism establishes in the interaction zone, very similar to the generation of tones in cavity flows and screeches in jets, which sustains the low-frequency large-scale shock unsteadiness. This mechanism was however refuted by \cite{wu2008analysis} in their DNS analysis for the case of interaction in the compression ramp. The study of correlation and time lags between the shock oscillations and the motion of the separation and reattachment locations suggested that the separation bubble drives the shock motion. A feedback loop is established among the separation bubble, the separated shear layer, and the shock system; disturbance of the mass balance between shear layer entrainment from the bubble and injection into the bubble at reattachment was suggested as the mechanism driving the low-frequency shock motion.

\begin{figure*}[hbt!]
\centering
\includegraphics[width=0.95\textwidth]{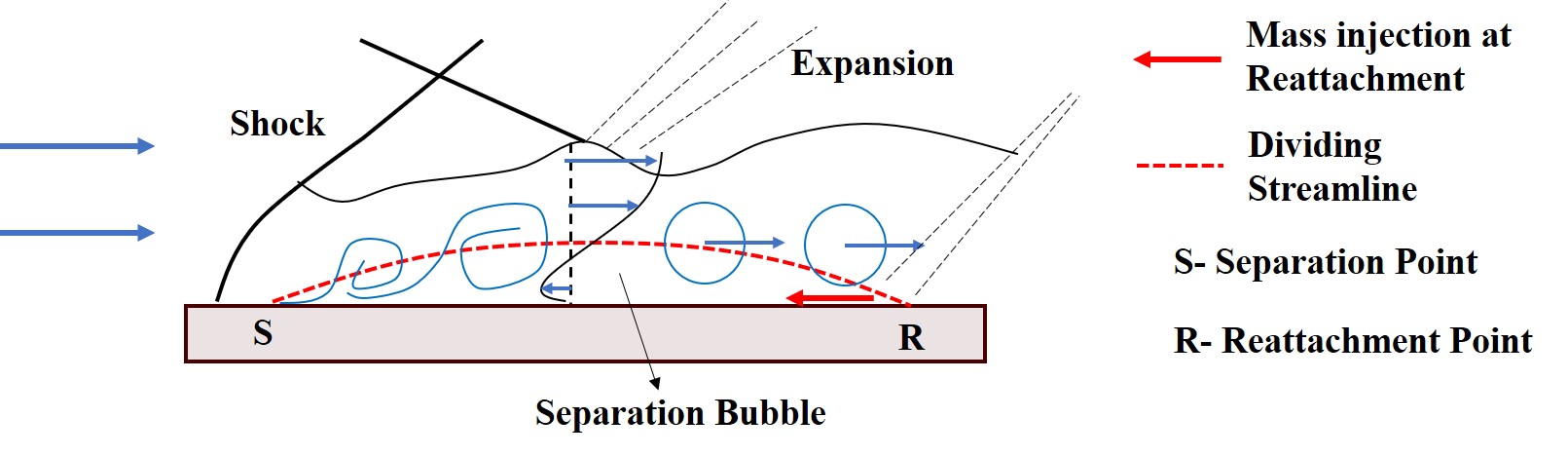}
\caption{Schematic of the separated flow field downstream of reflected shock in nominally 2-dimensional SBLI (adapted from \cite{piponniau2009simple})
}
\label{figure1:breathing-mechanism-piponniau-2009}
\end{figure*}

A simple model for the low-frequency unsteadiness of shock-induced separation, based on the relationship between successive contraction-dilatation of the separation bubble and the mass imbalance between shear layer entrainment and injection at reattachment, was proposed by \cite{piponniau2009simple}. Figure \ref{figure1:breathing-mechanism-piponniau-2009} shows a schematic of the shock-induced separated flow field in nominally 2-dimensional interactions. 

During the phases when the bubble is large, excess mass gets entrained through the dividing streamline in the relatively longer shear layer than the mass injection at reattachment. This drains the bubble, eventually causing it to collapse and shrink. As the bubble shrinks, the mass entrained in the shear layer reduces, and eventually excess mass gets injected at the reattachment, which tends to enlarge the bubble. The time scales estimated based on this `bubble breathing' mechanism were consistent with the experimentally observed shock oscillation Strouhal numbers. The DNS analysis by \cite{priebe2012low} also asserts the dependence of shock motion on the entrainment imbalances and the associated flapping of the separated shear layer, suggesting an inherent instability of the downstream separated flow. The review by \cite{clemens2014low} summarizes the present consensus on the shear layer entrainment-recharge mechanism behind the low-frequency dynamics. While spanwise wrinkling can be associated with the incoming turbulence \citep{ganapathisubramani2009low}, recent DNS studies by \cite{zhang2025analysis} established that the inherent instabilities associated with the low-frequency motion can also result in large-scale spanwise structures. \cite{clemens2009shock} and \cite{souverein2010effect} suggest that both upstream and downstream mechanisms jointly affect the interactions; however, the relative dominance of a particular mechanism depends on the interaction strength. Weak interactions without separation are dominated by the structure of upstream turbulent flow; however, for the cases of severe interactions with large separated regions, the downstream mechanisms are dominant.

While the unsteadiness associated with 2-D interactions has been widely investigated and resolved, the unsteadiness in 3-dimensional SBLI remains unclear for several 3-dimensional configurations. Due to the myriad 3-D configurations and qualitative differences in the flow field, achieving universality across them proved to be difficult. Among the 3-D configurations, swept shock interactions such as those due to sharp fins \citep{arora2018flowfield, arora2019unsteady} and swept compression ramps \citep{vanstone2019proper, adler2019flow, adler2020dynamics} have received greater attention recently. Unlike 2-D interactions, which have a closed separation (bubble) with spanwise homogeneity, 3-D interactions involve open separation, which, in the case of swept ramps/sharp fins, can exhibit quasi-cylindrical or quasi-conical symmetries. Due to these qualitative differences in the structure of the interaction, \cite{gaitonde2023dynamics} argued that the mechanism responsible for the low-frequency unsteadiness in two-dimensional interactions, viz., the bubble breathing, may not be present for three-dimensional interactions. Rather, in swept ramp-induced three-dimensional interactions, the shock motions were reported as shear-layer-induced events with mid-frequencies, i.e., $St_{\delta} \sim 0.1$. Apart from swept ramp interactions, the unsteady features in other 3-D interactions, particularly bow shock-induced interactions due to protuberances such as blunt fins, are not explored extensively. The bow shock-induced SBLI presents a curious case, in which, despite being a three-dimensional open separation, the separated flow is observed to exhibit low-frequency shock oscillations \citep{brusniak1994physics, ramachandra2023study}, similar to those observed in 2-D SBLI, and in contrast to the mid-frequency small amplitude oscillations observed in swept shock interactions.

In bow-shock induced separation, mean flow features have many qualitative differences with the 2-D SBLI. In 2-D interactions, the wall experiences a gradual pressure rise from interaction onset, followed by a plateau pressure in the separated region, and then another pressure rise due to flow reattachment. However, in bow-shock STBLI, the pressure distribution is qualitatively similar to two-dimensional interactions until some distance downstream of separation, with a wall pressure plateau after the pressure rise through separation, after which the pressure dips to a local minimum, before rising sharply to high peak values near the wall-protuberance junction \citep{dolling1982blunt, lindorfer2020scaling, bhardwaj2022scaling}. This low-pressure region is a distinctive feature of bow shock-induced interaction and is attributed to the local supersonic flow linked to the presence of a horseshoe vortex around the protuberance \citep{voitenko1966supersonic}. The dynamics of the horseshoe vortex, essentially a three-dimensional mechanism, can play a significant role in the overall unsteady dynamics of the SBLI, though this has not been addressed in the literature.

Concerning unsteady aspects in bow-shock induced STBLI, interesting observations were made in the experimental work of \cite{brusniak1994physics} on blunt fin-induced separation. Detailed unsteady wall pressure measurements were obtained along the spanwise centerline, upstream of the blunt fin in a Mach 5 flow on wind tunnel wall. The spectra in the intermittent zone (neighborhood of the shock foot/separation) were dominated by low frequencies, and those in the downstream separated flow were dominated by mid frequencies, which is strikingly similar to observations in nominally 2-D interactions exhibiting bubble breathing. However, marked differences with 2-D interactions are observed in the correlations between separation shock and various locations downstream. In 2-D interactions, the pressure fluctuations in the intermittent region are reported to exhibit a strong negative correlation with almost every other location inside the bubble (plateau pressure) and the reattachment. In contrast, for the bow shock-induced separation, the sense and the magnitude of correlation of pressure oscillations in the intermittent zone with those at downstream locations depended on the specific location. Strong negative correlations were observed with the pressure fluctuations in the plateau pressure zone as well as in the locations where the mean pressure decreases downstream up to the location of minimum mean pressure. Positive correlations were observed with the locations in the region of pressure rise from the minimum pressure point to the peak pressure point in the vicinity of fin-wall junction. The fluctuations at peak pressure point exhibited a negative correlation with the fluctuations in the intermittent zone. The magnitudes of correlation were relatively lower for downstream locations than those at the plateau pressure zone. At the locations between the zones exhibiting correlation peaks of opposite sense with the fluctuations at intermittent zone, both positive and negative peaks in correlation were observed. These observed trends in the correlations with shock foot pressure fluctuations were clearly linked to the distinct surface pressure distribution.

\begin{figure*}[hbt!]
\centering
\includegraphics[width=0.95\textwidth]{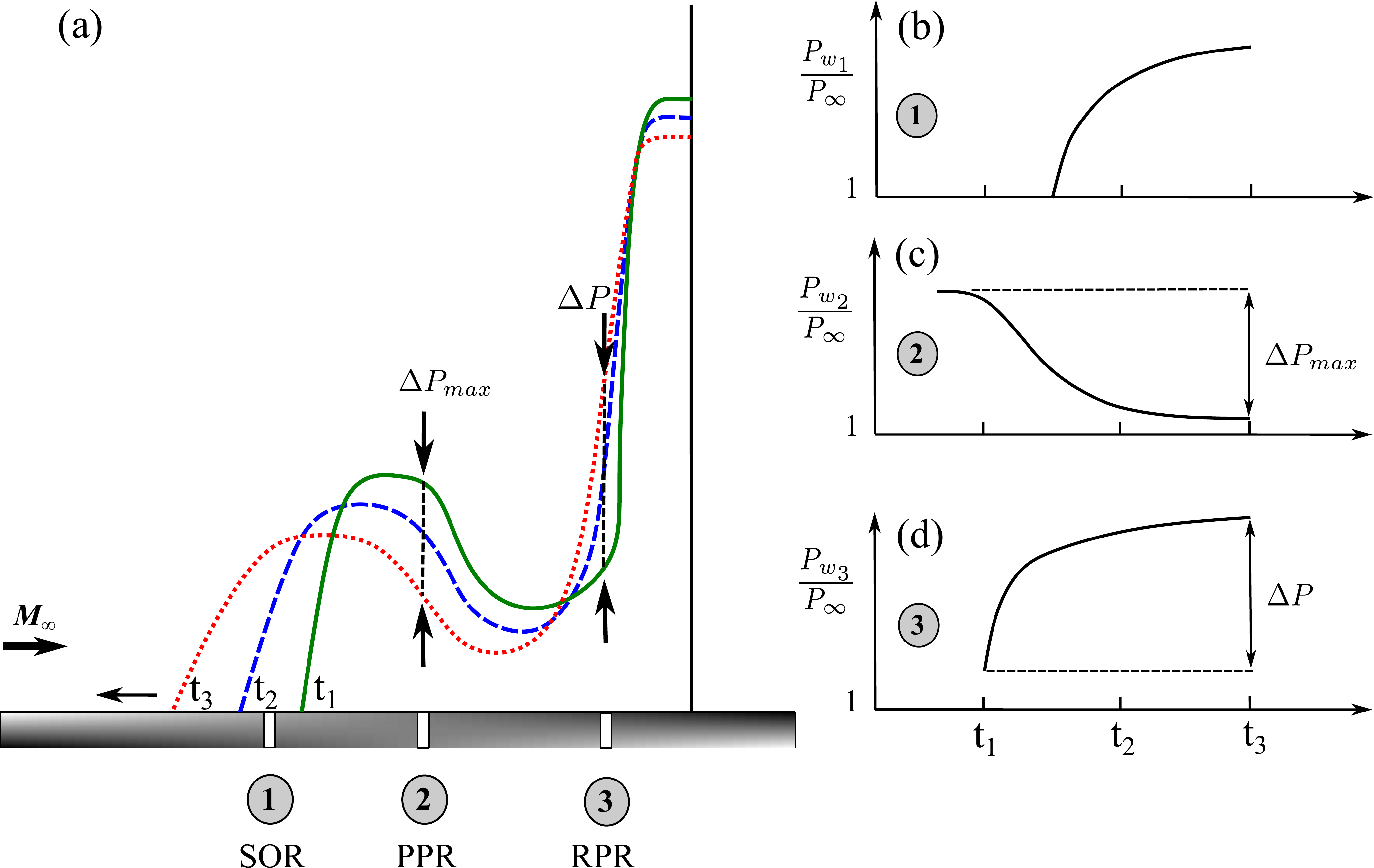}
\caption{(a) Ensemble-averaged pressure distribution; pressure-time variation for upstream shock foot movement at (b) shock oscillation region (SOR); (c) plateau pressure region (PPR); (d) rising pressure region (RPR) (adapted from \cite{brusniak1994physics})
}
\label{figure2:wallpressure-variation-Brusniak-1994}
\end{figure*}

The wall pressures at a particular location, measured at all instances when the shock foot is at a given streamwise location, is averaged to give the ensemble average at the particular location corresponding to the given shock foot location. Figure \ref{figure2:wallpressure-variation-Brusniak-1994}(a) schematically shows the ensemble-averaged pressure distribution along the centerline for three progressive time instances as the shock moves upstream, viz., $t_1$, $t_2$ and $t_3$. The pressure-time variation at three probe locations: location 1 in the intermittent/shock oscillation region (SOR), location 2 in the plateau pressure region (PPR), and location 3 in the rising pressure region (RPR) between the local minimum and the peak pressure near the base of the cylinder- are traced in Figure \ref{figure2:wallpressure-variation-Brusniak-1994}(b),(c) and (d), respectively, based on the behaviour of the ensemble averaged  from $t_1$ to $t_3$. Based on the observed trends, it can be inferred that the low-frequency wall pressure fluctuations at mean shock-foot must have a strong negative correlation with the fluctuations in the plateau pressure region due to both the stretching and flattening of the pressure profile, whereas shock-foot pressure fluctuations should have a positive correlation with those at location 3 due to the stretching of the profile. The flattening of peak pressure must also result in a negative correlation between fluctuations at mean shock and peak pressure location near cylinder base. The stretching and flattening of the ensemble-averaged pressure distribution as the shock foot moves upstream (and vice versa for downstream motion), thus explains the observed trends in the correlations in wall pressure fluctuations. Although this analysis explains the observed correlations in unsteady wall pressure at spanwise centerline, the mechanism sustaining the low-frequency unsteadiness in the bow shock-induced open separation remains unaddressed. 

The experimental investigations on transitional or turbulent SBLI over blunt fin \citep{combs2019unsteady, lindorfer2020scaling, hoffman2022modal} also relied on centerline unsteady wall pressure measurements or time series of centerline shock foot position from schlieren visualizations. High-fidelity computational studies on these interactions are also limited. These interactions pose significant computational challenges in characterizing shock unsteadiness. RANS models lack the ability to capture low-frequency shock oscillations due to rapid distortion of the flow field \citep{marshall1992computation,bhardwaj2022scaling}, whereas high-fidelity simulations (DNS/LES), although capable of resolving them, require prohibitively expensive grids and longer computational times. This necessitates the use of hybrid methods, such as detached eddy simulations (DES), especially for flows at high Reynolds numbers from an engineering perspective \citep{spalart1997comments, spalart2006new}. In these lines, a numerical work by \cite{ngoh2022detached} employed DES for blunt fin-induced interactions. However, the reported spectra and correlations were only along the centerline; although the pressure oscillations at other spanwise planes could be accessible from the simulations, lack of experimental data for validation may be a reason for not looking into the spanwise aspects of oscillations. As mentioned before, the spectra along the spanwise centerline exhibit behavior similar to that in 2-D SBLI, which is perplexing, given that there is no trapped mass inside the separated region. It is not appropriate to argue in favor of the breathing mechanism merely on the basis of frequency ranges; the horseshoe vortex and three-dimensional relieving effects may have some contribution. Simultaneous measurements at various spanwise locations and exploration of space-time correlations are imperative. Essentially, a ‘three-dimensional’ view of the unsteady flow field needs to be carefully looked at to understand the mechanism. Only recently, a numerical work \citep{lindorfer2025characterizing} has attempted to address this through LES on blunt fin-induced separation, reporting a limited observation that the intermittent length and the (low) frequency content were nearly the same across the span in the regions where the flow separation was present. 

In this context, the present work investigates the bow shock-induced interaction due to protuberances through experimental and computational approaches to resolve the mechanism. An emphasis is given to the correlations and spectra at different spanwise locations, in addition to the centerline span. Wind tunnel experiments are conducted at a Mach number of 2.87 and a Reynolds number of $4.7 \times 10^{7} \,\ m^{-1}$, with protuberances of different shapes and sizes mounted on the tunnel wall. The unsteadiness was characterized using time-resolved schlieren visualizations and unsteady pressure measurements at various streamwise and spanwise locations, from which the scaling for frequency and amplitude of shock oscillations are obtained. Further, the interaction is investigated numerically using an enhanced version of DES, `adaptive DES (ADES)', which strikes a balance between computational requirements and the desired level of accuracy. The simulations are conducted for the case of a cubical (square-faced) protuberance, whose results are validated against the experimental data. Computations are not intended to be used for a parametric study, but only for a finer resolution of flow physics. Therefore, we chose to perform simulations only with the cubical protuberance. Based on the insights from the experiments and computations, the mechanism sustaining the low-frequency shock motion is explored. The present article is organized as follows. Details of the experimental methodology and important findings obtained for various protuberances are provided in Section \ref{sec2:expt-methodology}. Section \ref{sec3:comp-methodology} presents the computational methodology, including the flow configuration and the numerical methods used for the cubical protuberance case. Section \ref{Sec4:comp-observations} is devoted to computational results for cubical protuberances, including validation studies, spectral analyses, and insights from two- and three-dimensional POD analyses, leading to a final discussion on the governing mechanism for low-frequency shock oscillations.

\section{Experimental Methodology and Observations}
\label{sec2:expt-methodology}
\begin{figure*}[!htb]
    \centering
    \includegraphics[width=0.95\textwidth]{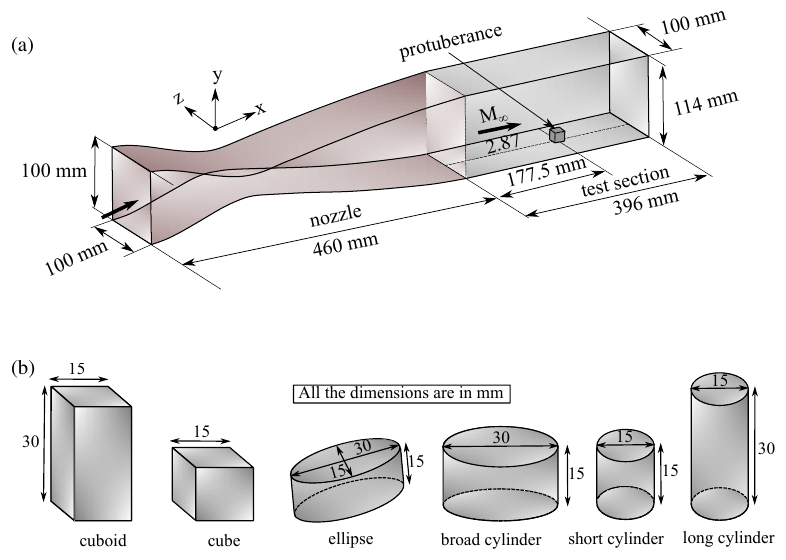}
    \caption{Schematic of (a) experimental setup. (b) different protuberances considered in the experiments.} 
    \label{figure3:expt-setup-protuberance-geom}

\end{figure*}

The experimental investigation is conducted in the supersonic blowdown wind tunnel at the Gas Dynamics Laboratory at IIT Madras. The test section of the tunnel shown in figure \ref{figure3:expt-setup-protuberance-geom}(a), having a rectangular cross section of 100 mm $\times$ 114 mm and a length of 396 mm, is supplied with compressed dry air at a stagnation pressure of 6 bar. The stagnation temperature is the room temperature of 300 K. A contoured nozzle expands the flow from settling chamber to a freestream Mach number of 2.87 in the test section; the freestream pressure ($p_{\infty}$) and temperature ($T_{\infty}$) are 19200 Pa and 113 K, respectively. More details can be found in the works of \cite{bhardwaj2021} and \cite{ramachandra2023}. In the current study, protuberances of different shapes and sizes (as shown in figure \ref{figure3:expt-setup-protuberance-geom}(b)) are employed to understand the universal features of protuberance-induced separations. The protuberances are mounted on the bottom wall of the test section at the spanwise centerline, the span of the test section being 100 mm. The square-faced (cubical) and the rectangle-faced (cuboidal) protuberances are positioned with one of their faces normal to the incoming flow. The elliptical protuberance is mounted with its major axis aligned with the streamwise direction, whereas cylindrical protuberances are fixed to the bottom plate with one of their flat sides. All the protuberances are positioned so that their mounting centers are 177.5 mm from the nozzle exit. The boundary layer thickness at the location of the protuberance is measured to be 7 mm \citep{bhardwaj2021}.

        %% ----------------------------------------------------------------------------------------------------

The oil flow visualization was used to obtain surface streakline patterns that highlight mean interaction features, including the mean separation line. The oil mixture prepared from Titanium dioxide (Ti$O_{2}$) powder, Oleic acid, and SAE 30 oil was spattered on the base plate before the run. The centerline separation lengths ($L_{sep}$), defined as the streamwise distance between the protuberance leading edge and the separation line at spanwise center, are then measured from the images. The Z-type schlieren imaging technique was employed for the time-resolved visualization of the interaction using a high-speed camera (Photron FASTCAM SA4 Model 500K-M1). The collimated light is directed normal to the optical windows at spanwise ends of the tunnel, and thus the schlieren images present the density gradient information (line of sight integrated) in the planes normal to the spanwise direction. The snapshots were obtained at 30000 frames per second (fps), with a shutter speed of 1/35000 seconds and a spatial resolution of $384 \times 288$ pixels. A total of 8000 continuous snapshots corresponding to a time of 0.27 s with a time resolution of 33$\upmu$s were used for all the analyses presented in this work. Time-averaged surface pressures were measured through static pressure taps on the bottom wall and the front face of the protuberance. Scanivalve DSA 3217/16Px with a rated pressure of 50 psi was used to carry out these measurements. The device can measure pressures from 16 ports simultaneously. With a good spatial resolution of mean wall pressures, through taps at numerous streamwise and spanwise locations, various pressure zones in the interaction region could be clearly mapped. The unsteady surface pressure signals were extracted in critical interaction regions using unsteady pressure transducers, such as Kulite (XCQ-062) and Endevco (8530C). Three Kulite sensors with a rated pressure of 1.7 bar (abs) and a sensing head diameter of 1.7mm were used simultaneously per experimental run. For the experiments measuring pressure at reattachment, an Endevco sensor (8530C) with a relatively large sensing area (outer diameter of 3.86 mm) was used. The signals were acquired at 500 kHz for 2 seconds using the NI-cDAQ 9185 and NI-9222 data acquisition module.

\subsection{Data processing techniques}
\subsubsection{power spectral density}
\label{sec2.1.1:power-spectral-density}
Power spectral density (PSD), which represents the power or strength of the signal in the frequency domain, is calculated for all the signals using \textit{Welch}'s windowing method \citep{welch1967use}. A MATLAB inbuilt function \textit{pwelch} is utilized for this. \textit{Hamming} windows having a size of 25000 samples with 50\% overlap were used for the spectral estimations. The PSDs were premultiplied by frequency and normalized by the corresponding variance of the signal and are plotted against the Strouhal number based on boundary layer thickness, $St_{\delta}$. 

\subsubsection{cross correlation}
\label{sec2.1.2:cross-correlation}
The statistical relations between signals at different locations were analyzed through correlation and coherence coefficients. Correlation between two signals measures the similarity of the first signal with the second at various time lags ($\tau$). For instance, consider two signals $x(t)$ and $y(t)$, then the correlation function ($C_{xy}$) and the correlation coefficient ($R_{xy}$) are defined as follows.

\begin{equation}
    C_{xy}(\tau) = \frac{1}{N} \sum_{t=1}^{N-\tau} (x(t)-\overline{x}) (y(t+\tau) - \overline{y})
\end{equation}

\begin{equation}
    R_{xy}(\tau) = \frac{C_{xy}(\tau)}{\sqrt{C_{xx}(0)} \sqrt{C_{yy}(0)}}
\end{equation}

where $\overline{x}, \overline{y}$ are mean values of the corresponding signals. The magnitude of $R_{xy}(\tau)$ tells the level of similarity between the signals at the corresponding time lag ($\tau$), and sign signifies if they are `similar' or `anti-similar'. For positive values of $\tau$, $x(t)$ leads $y(t)$, and vice versa for negative $\tau$.

\subsubsection{coherence}
\label{sec2.1.3:coherence}
The coherence between the signals conveys how the two signals are linearly related to each other at a given frequency. It is estimated using magnitude-squared coherence, $\gamma^2_{xy}(f)$ defined as follows:

\begin{equation}
    \gamma^2_{xy}(f) = \frac{|P_{xy}(f)|^2}{P_{xx}(f) P_{yy}(f)}
\end{equation}

where $P_{xx}$, $P_{yy}$ are PSD of signals $x(t)$ and $y(t)$, respectively. $P_{xy}$ represents cross PSD of $x(t)$ and $y(t)$. $\gamma^2_{xy}$ is always positive and ranges from 0 to 1. Higher the coherence value, higher the linear relation between signals at the corresponding frequency. 

\subsubsection{proper orthogonal decomposition}
\label{sec2.1.4:snapshot-pod}
Proper orthogonal decomposition (POD) was employed to identify the dominant flow structures and the associated frequencies. POD is a method of decomposing the spatio-temporal flow field U(x,t) into a set of orthogonal spatial modes and temporal coefficients, whose sum reconstructs the original flow field. 

\begin{equation}
    U(x,t) = \sum_{k=1}^{\infty} a_{k}(t) \ \ \Phi_{k}(x)
\end{equation}

% \begin{equation}
%     A(x,t) = \sum_{n=1}^{\infty} U_{n}(x) \ \ v_{n}(t)
% \end{equation}

where $\Phi_{k}(x)$ represents $k^{th}$ spatial mode and $a_{k}(t)$ is corresponding time coefficient. 
These modes represent correlated spatial structures in the flow field. In POD of discrete time data, these modes are obtained by the singular value decomposition (SVD) of a matrix formed by the sequential arrangement of the vectors containing quantities of instantaneous spatial field. By arranging the singular values in decreasing orders of their magnitude, the corresponding modes too are arranged in the order of their energy. This is computed effectively using the snapshot POD algorithm \citep{sirovich1987turbulence}.

\begin{figure*}[hbt!]
\centering
\includegraphics[width=0.98\textwidth]{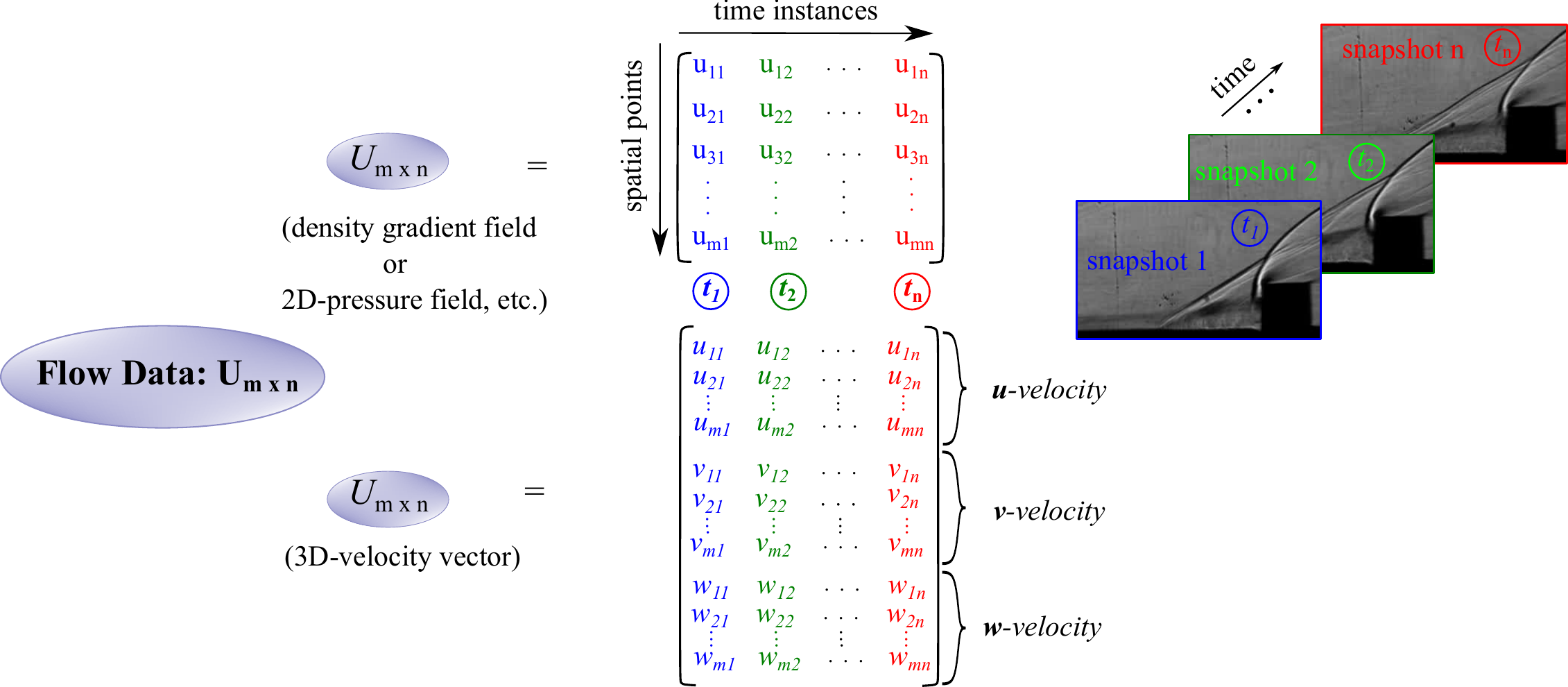}
\caption{Fluctuating data matrix construction in snapshot POD}
\label{figure2:pod-data-matrix}
\end{figure*}

In this method, the SVD is reduced to the eigenvalue problem of the correlation matrix $C = U^TU$. Here each column vector of U contains the discrete spatial variation of fluctuations in quantities of interest at a given time instance, and column vectors are arranged in increasing order of time, as shown in figure \ref{figure2:pod-data-matrix}. The fluctuations are essentially the instantaneous local values subtracted by the time-averaged local value of the quantity of interest. In the case of experimental schlieren images, the fluctuations in intensities in all the $m$ pixels in an instantaneous schlieren image constitute a particular column; and the columns corresponding to all $n$ instances are arranged sequentially to form the $m \times n$ matrix U. In the case of numerical data, two different planar data are analysed using POD\--- the density fluctuations at the spanwise centered wall-normal plane (which can be compared with the POD of experimental schlieren), and the pressure fluctuations at the bottom wall. In each of these cases the columns of the U matrix are made of instantaneous fluctuations of the respective quantities at all the grid points in the respective plane. The three-component velocity data from computations is also analyzed using 3-D POD, as illustrated in figure \ref{figure2:pod-data-matrix}. The eigenvalues and eigenvectors of the correlation matrix are obtained from the following eigenvalue problem.
\begin{equation}
    C A = \Lambda A
\end{equation}

where eigenvectors $A_{n \times n} = [a_1 \ \  a_2 \ \ a_3 \ \  ... \ \ a_n] $ represents the matrix of temporal coefficients, and $\Lambda$ represent diagonal matrix of eigen values. Eigenvalues are sorted and arranged in descending order ($\lambda_{1} > \lambda_{2} > \lambda_{3} > \ ....\lambda_{n}$), and the corresponding eigenvectors are also arranged accordingly. Subsequently, the spatial modes ($\Phi$) are computed by projecting the data ($U$) onto the eigenvectors ($A$) computed from the eigenvalue decomposition of the covariance matrix ($C$).
\begin{equation}
    \Phi = U A
\end{equation}

The original flow field can be reconstructed using the first few dominant spatial modes and time coefficients, rather than the total number ($n$) of modes. Cumulative energy captured by the first `r’ number of modes, where $r < n$, gives us an estimate of how well the modes under consideration represent the flow field. As eigenvalues are arranged in decreasing order of their magnitudes, then the amount of energy captured by the first $r$ modes is given by: 
\begin{equation}
    E_{r} = \frac{\sum_{i=1}^{r} \lambda_{i}}{ \sum_{i=1}^{n} \lambda_{i}}
\end{equation}
Therefore, reduced-order modeling is performed to obtain the flow field using a few modes with high energy (information or variance) and corresponding time coefficients. The flow field reconstructed by the first $r$ modes is given by:

\begin{equation}
    \widetilde{U} = \widetilde{\Phi} \widetilde{A}^{T}
\end{equation}

where $\widetilde{\Phi}_{m \times r} = [\Phi_{1} \ \ \Phi_{2} \ \  \Phi_{3} \ \ ... \ \  \Phi_{r}]$ and $\widetilde{A}_{n \times r} = [a_{1} \ \ a_{2} \ \  a_{3} \ \ ... \ \  a_{r}]$. 

%%%%%%%%%%%%%%%%%%%%%%%%%%%%%%%%%%%%%%%%%%%%%%%%%%%%%%%%%%%%%%%%%%%%%
\subsection{Mean flow features}
\label{sec2.2:mean-features-experiments}
\begin{figure*}[!htb]
    \centering
    \includegraphics[width=0.85\textwidth]{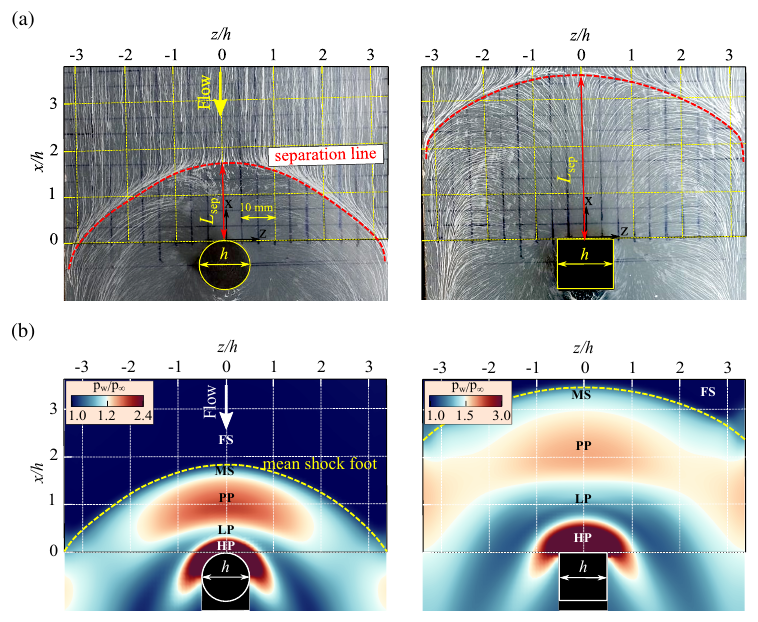}
    \caption{Mean flow fields of the interaction (a) oilflow streakline patterns (b) surface pressure from RANS simulations \citep{bhardwaj2022scaling}; FS - freestream, MS - mean shock foot, PP - plateau pressure zone, LP - low pressure zone, HP - high pressure zone}
    \label{figure3:streaklines-surface-pressure}    
    
\end{figure*}

The time-averaged view of the interaction is examined for different protuberances based on separated-flow topology and static pressure measurements. The inviscid bow shock in front of the protuberances interacts with the boundary layer, resulting in 3-dimensional separation. Protuberances with varied geometric features resulted in different extents of separated regions, owing to the differences in the inviscid bow shock shape, and thus the differences in spanwise shock-imposed adverse pressure gradient. Figure \ref{figure3:streaklines-surface-pressure}(a) shows the streakline patterns obtained from the oil flow visualizations, for short cylinder and cuboid, observed in the current investigation. The respective separation lines are also traced on the figures. Despite the differences in the separation extents, i.e, small separation for cylinder and large separation for cuboid, the interaction features are found to be qualitatively similar. This is further illustrated through the time-averaged pressure field (see figure \ref{figure3:streaklines-surface-pressure}(b)) from RANS simulations \citep{bhardwaj2022scaling}, whose values compare well with the surface pressures measured using the electronic pressure scanner at various distinct locations. Distinct zones can be identified based on the surface pressure distribution, including freestream (FS), intermittent zone from which the mean shock (MS) foot can be traced, plateau pressure (PP), low-pressure (LP), and high-pressure (HP) zones. Although the extents of these zones are dependent on the protuberance geometry, the features of the interaction are qualitatively similar for all the protuberances.    

%%%%% Oil flow viscualization and wall-pressure distirbution :: Short-cylinder, Cuboid %%%%%

The pressure variation along the spanwise centerline ($z/h = 0$) is qualitatively similar to the 2-D interactions in the upstream part of the separation zone. The pressure rises gradually across the separation shock foot, through the separation and reaches a plateau pressure. However, unlike the 2-D interactions in which the pressure rises monotonically after the plateau as the flow reattaches, in the present case, a fall in pressure is observed after the plateau, constituting the `low-pressure zone', before it rises again to very high pressures near the protuberance base. The low-pressure zone is caused by locally accelerating reverse flow above the wall (which can be supersonic), forming a horseshoe vortex that spirals out and relieves the flow in the spanwise direction \citep{hung1985simulation, bhardwaj2022scaling}. Based on the time-averaged characterization of the flow field, the locations for wall pressure measurements using fast response transducers was decided. The subsequent section discusses the characterization of the unsteady flow field. The experimental work by \cite{ramachandra2023study} had discussed some unsteady features for the case of square-faced/cubical protuberances, which are referred here to briefly describe the unsteady aspects, and to then establish the universal nature of those observations, applicable for various protuberance shapes.

%%%%%%%%%%%%%%%%%%%%%%%%%%%%%%%%%%%%%%%%%%%%%%%%
\subsection{Unsteady flow features}
\label{sec2.3:unsteady-features-experiments}
\begin{figure*}[!htb]
    \centering
    \includegraphics[width=0.40\textwidth]{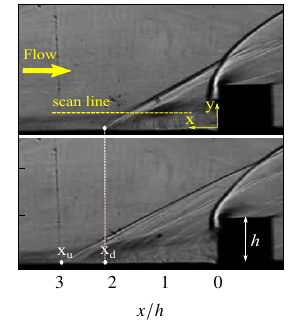}
    \caption{Instances of upstream and downstream-most shock positions. Separation shock (SS), reattachment shock (RS), and scan lines are also marked in the figure.}
    \label{figure4:schlieren-expt-upstream-downstream}    
    
\end{figure*}

\begin{figure*}[!htb]
    \centering
    \includegraphics[width=0.95\textwidth]{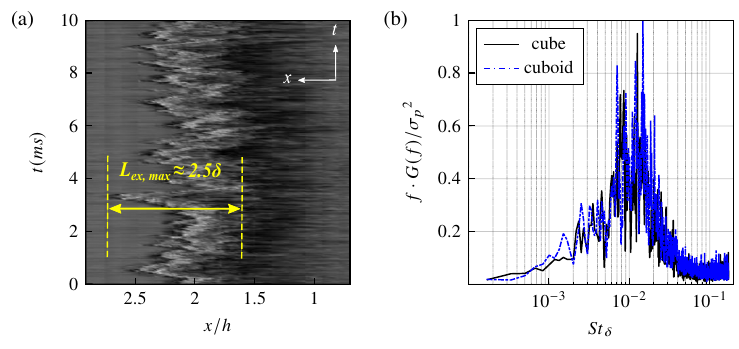}
    \caption{Separation shock unsteady characteristics along the scan line (a) space-time (x-t) variation  (b) power spectral density.}
    \label{figure5:x-t_scanline-experiments}    
    
\end{figure*}

\begin{figure}
    \centering
    \includegraphics[width=0.95\linewidth]{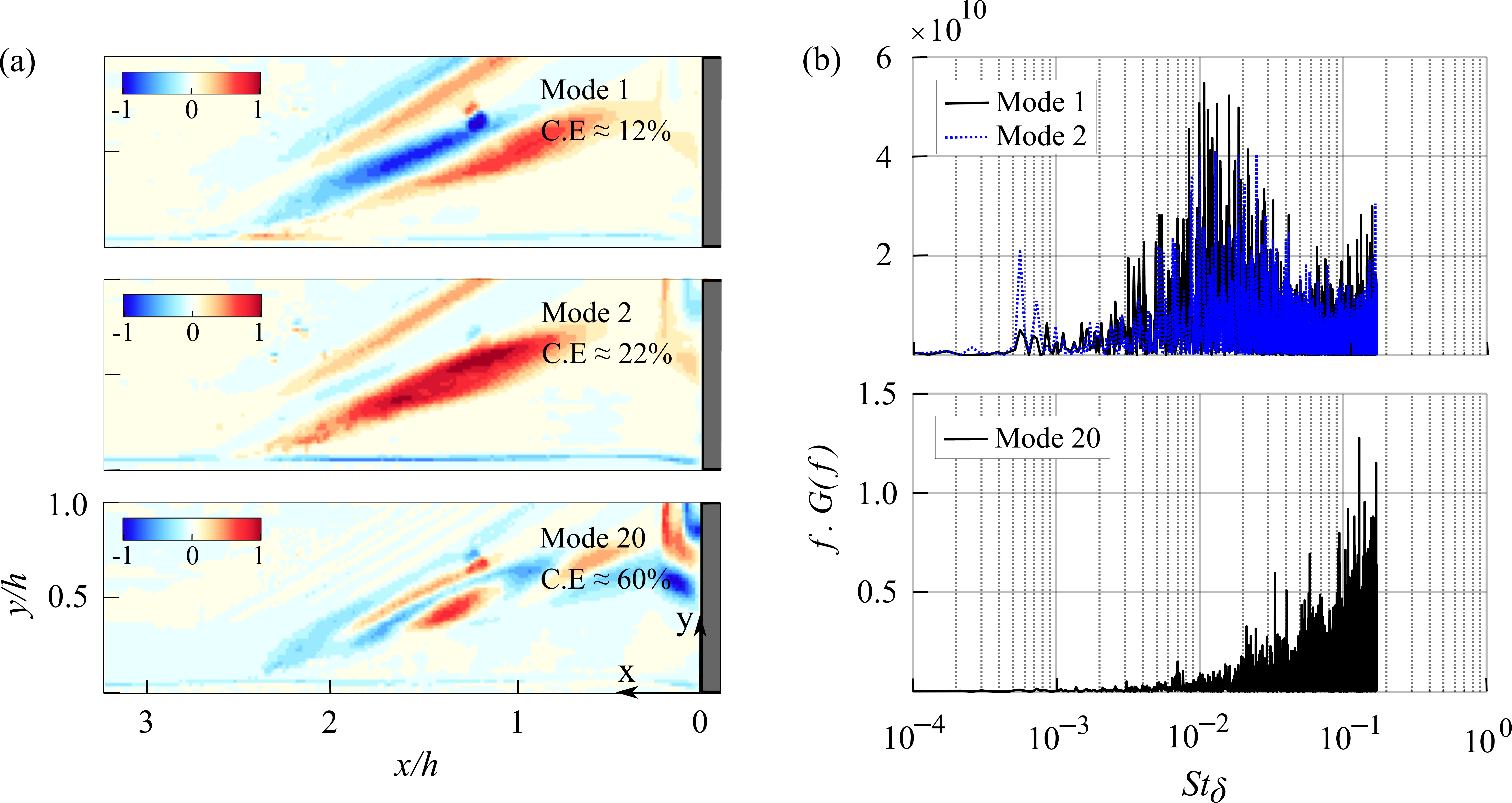}
    \caption{Centerline span POD of cubical protuberance (a) spatial modes. (b) PSD of corresponding temporal coefficients}
    \label{figure8:centerline-pod}
\end{figure}

The unsteady aspects of the interaction are characterized from the time-resolved schlieren images and the wall pressure signals from the transducers. The bow shock-induced separation results in a separation shock, whose foot is upstream of the separation line, and the separated flow at the spanwise centerline reattaches on the protuberance face, resulting in the reattachment bow shock. Figure \ref{figure4:schlieren-expt-upstream-downstream} presents two instantaneous schlieren snapshots when the shock foot is at its upstream and downstream-most positions, marked as $X_u$ and $X_d$, respectively, during an oscillation cycle. The separation and reattachment shocks are marked as SS and RS, respectively. It is apparent from the snapshots that for the given cycle of shock motion, the intermittent region spans over a distance comparable to $h$. However, the intermittent length can vary across different shock cycles and can be visualized using a space-time (x-t) plot of shock movements. This is obtained by scanning the pixel intensities along a scan line parallel to the bottom wall, which in this case is at $0.92\delta$ from the bottom wall. From the x-t variation (see figure \ref{figure5:x-t_scanline-experiments}(a)), the maximum shock excursion is estimated to be around $2.5\delta$ or $1.17 h$. The shock foot position as a function of time was also extracted from the scan line intensity values by tracing the jump in intensity. Figure \ref{figure5:x-t_scanline-experiments}(b) shows the PSD of shock-foot motion, from which it is evident that the low-frequencies\--- $St_{\delta} \sim 0.01$\--- are dominant, i.e., the shock foot exhibits low-frequency oscillations. The spectra of shock motion with other protuberances too exhibit the dominance of similar frequency ranges, as illustrated with the PSD for the cuboid case in figure \ref{figure5:x-t_scanline-experiments}(b), though the extents of the separated region are different for different protuberances. From the POD analysis based on schlieren snapshots, as shown in figure \ref{figure8:centerline-pod} for the cubical protuberance, the first two energetically dominant modes represent low-frequency shock oscillations ($St_{\delta} \sim 0.01$) along with associated shear layer movements, accounting for approximately 22\% of the energy. The higher modes show mid-frequency shear layer events. For example, in mode 20 exhibiting large scale structures in the shear layer, mid-frequencies ($St_{\delta} \sim 0.1$) are seen to be dominant. Further low-energetic modes are characterized by small-scale structures. Thus, while the region in the vicinity of the shock-foot is expected to be dominated by low-frequency oscillations in flow variables, within the separated region, mid-frequency shear layer oscillations are also expected to be felt.  

%%%%%%%%%%%%%%%%%%%%% Centerline-shock-movemnts-cube ::  PSD:: Cube-Cuboid %%%%%%%%%%%%%%%%%%%

%%%%%%%%%%%%%%%%%%%

%\FloatBarrier
%%%%%%%%%%%%
%%%%%%%%%%%%

\begin{figure}
    \centering
    \includegraphics[width=0.6\textwidth]{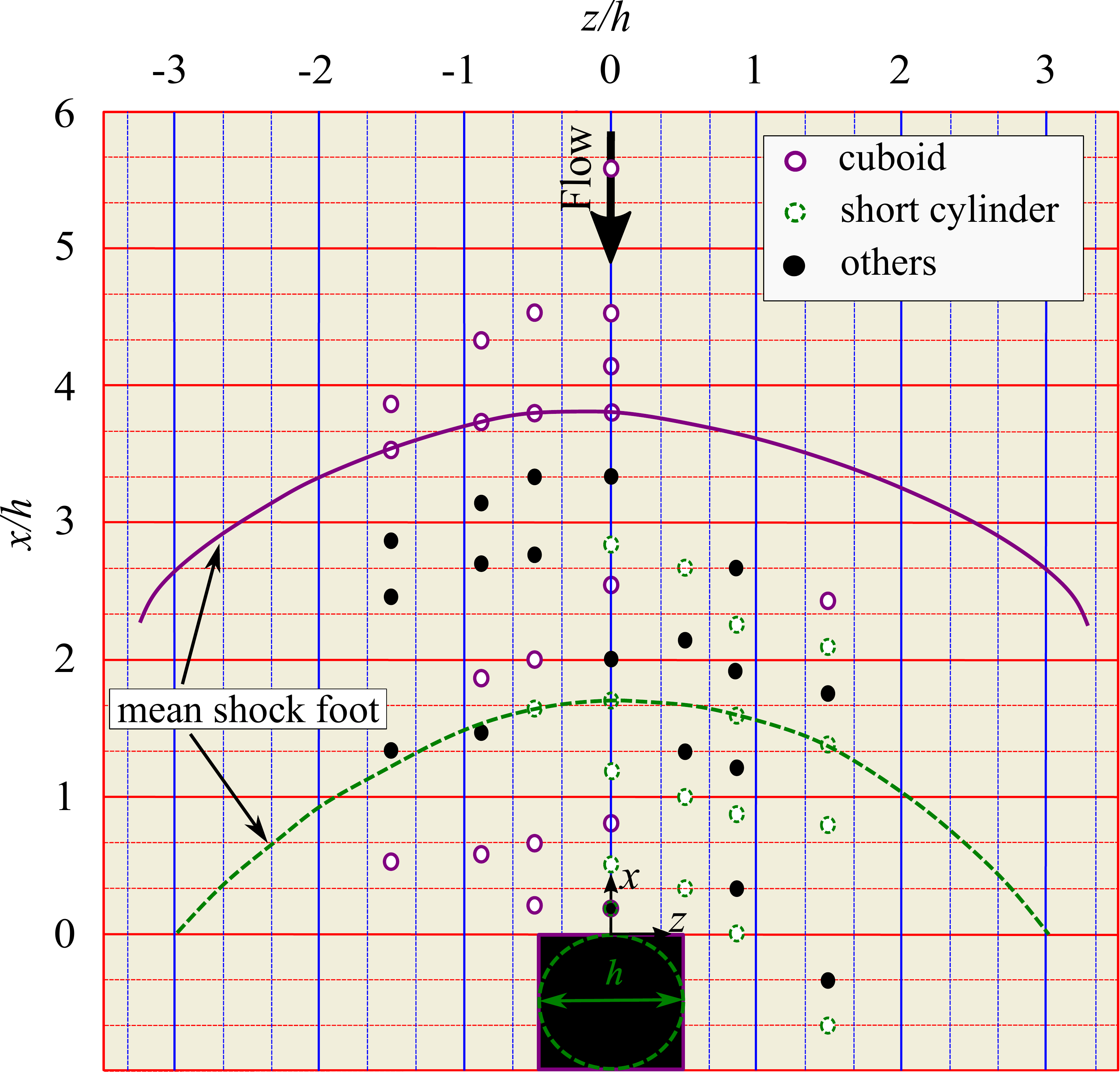}
    \caption{Schematic of transducer locations strategized for the pressure measurements}
    \label{figure6:probe-locations-experiments}
\end{figure}

Having characterized the time-averaged picture and noting the different pressure zones, the transducers are strategically placed to resolve the spatio-temporal unsteady aspects. To characterize the 3-dimensional aspects and the mechanism, it is imperative to look at the relationship between wall pressure oscillations at different locations, not just along the spanwise centerline (as presented by \cite{brusniak1994physics}), but also at various spanwise locations. Figure \ref{figure6:probe-locations-experiments} shows a schematic of the transducer placement on the wall. The locations used for two of the protuberances, viz. cuboid and short cylinder, are particularly highlighted to illustrate the strategy of placement of the transducers in various spanwise and streamwise locations. For instance, the mean shock foot for those protuberances are marked, from which the transducer placements at different spanwise locations along the mean shock foot can be noted; similarly in other pressure zones too, transducers were placed at different spanwise locations for the respective case. 

\begin{figure*}[!htb]
    \centering
    \includegraphics[width=0.45\textwidth]{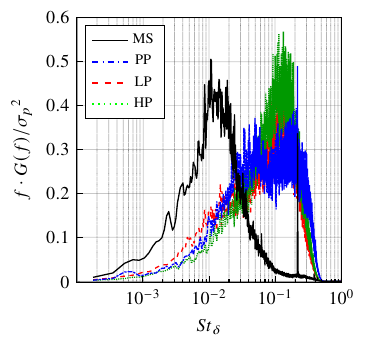}
    \caption{Power spectral densities of flush-mounted probes along the spanwise centerline for cubical protuberance.}
    \label{figure7:centerline-probes-psd-flush-mounting-experiments}    

\end{figure*}

\begin{figure*}[!htb]
    \centering
    \includegraphics[width=0.95\textwidth]{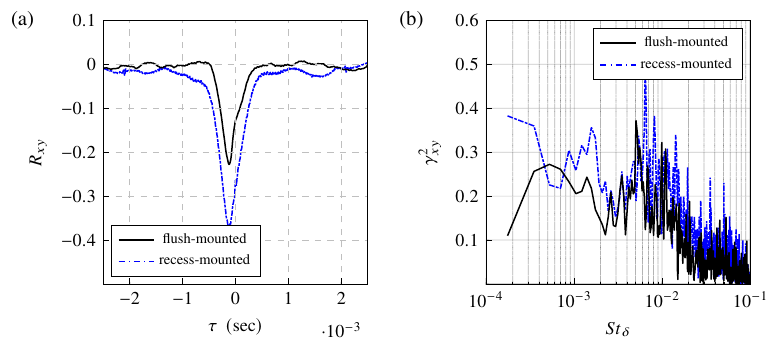}
    \caption{Statistical relations of signals between mean shock and low-pressure zone for cubical protuberance (a) cross correlation. (b) coherence.}
    \label{figure8:centerline-statistical relation-MS-LP-experiments}    
    
\end{figure*}

As noted by \cite{brusniak1994physics}, the correlations in surface pressure are highly sensitive to changes in the location, and it is desirable to have transducers with small sensing areas. The fast response Kulite (XCQ-062) sensors used in the current study have a relatively small sensing area (diameter $\sim$ 1.7 mm). However, the non-threaded arrangement of the sensors led to issues with flush mounting to the wall, due to concerns such as the limited number (three) of available transducers, which had to be moved to different locations for different experiments, and the safety of the transducers. Therefore, a recess/cavity mounting strategy was devised following the works of \cite{verma2019control, verma2025normal}. With this strategy, the diameter of the sensing surface is further reduced to 0.5 mm, which is advantageous from the perspective of point measurements and consequently, for two-point correlations. Details about the recess mounting strategy can be found in \cite{ramachandra2023}. Recess mounting can cut off higher frequencies but captures the frequencies associated with shock boundary layer interactions (\cite{verma2019control, verma2025normal}). Therefore, to present the complete spectra of wall pressure fluctuations across the different zones of interest, measurements with flush-mounted Endevco transducers (8530C) with larger sensing areas (diameter $\sim$ 3.86 mm) are used. Figure \ref{figure7:centerline-probes-psd-flush-mounting-experiments} shows the premultipled PSD of the wall pressures measured using flush-mounted Endevco transducers at locations in different zones along the spanwise centerline, for the case of the cubical protuberance. The pressure fluctuations in the intermittent zone (at MS) are dominated by low frequencies, whereas at other locations within the separated flow, mid-frequencies are dominant, though there is considerable energy in the low-frequency range corresponding to shock oscillations. However, the trends in correlations were similar for both flush and recess-mounted cases, despite the fact that higher frequencies are cut off by recess mounting. In fact, the magnitudes of the correlations are observed to be considerably enhanced in recess-mounted cases, as illustrated in figure \ref{figure8:centerline-statistical relation-MS-LP-experiments}(a), showing the correlations between MS and LP locations for the case of cubical protuberance obtained with different mounting strategies. This is because of high coherence in the low frequency range as given in figure \ref{figure8:centerline-statistical relation-MS-LP-experiments}(b). Therefore, since our primary interest is in correlations affecting low-frequency motion, we shall use data from the recess mounting. 

% \begin{figure}
%     \centering
%     \includegraphics[width=0.95\linewidth]{Experimental-results/Kulite-recess-mounting.pdf}
%     \caption{Caption}
%     \label{fig:placeholder}
% \end{figure}

%%%%%%%%%%%%%%%%%%%%% Centerline PSD:: Endevco %%%%%%%%%%%%%%%%%%%
%%%%%%%%%%%%%%%%%%%%%%%%%%%%%%%%%%%%%%%%%%%%%%%%%%%%%%%%%%%%%%%%

%%%%%% Correlation & Coherence:: MEAN SHOCK && LOE PRESSURE %%%%%%
%% Flush and Recess Mounting

\subsection{Correlation and coherence studies of shock oscillations}
\label{sec2.4:corr-coher-experiments}
\begin{figure*}[!htb]
    \centering
    \includegraphics[width=0.45\textwidth]{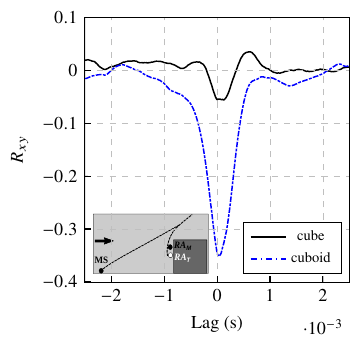}
    \caption{Correlation between separation shock and reattachment locations; MS - mean shock foot, $RA_M$ - measured reattachment location, $RA_T$ - true reattachment location.}
    \label{figure9:correlation-sep-reat-experiments}    
    
\end{figure*}

Regarding the centerline correlations among various zones of the interaction, the mean shock foot exhibits significant correlations with downstream regions, and the sense of correlation changes as the distance increases in the downstream direction towards the protuberance; a strong negative correlation is observed with the plateau pressure region, and a positive correlation with the low-pressure region, followed by a correlation switch to negative in the high-pressure region. Such trends in the correlation were already discussed for the cube in \cite{ramachandra2023study}, and are qualitatively similar for other protuberances as well, and are similar to those reported by \cite{brusniak1994physics}, which is discussed in detail in the introduction. A considerable negative correlation was also observed between separation shock foot (MS) and reattachment ($RA_{M}$) in the cuboid case as shown in figure \ref{figure9:correlation-sep-reat-experiments}. However, the same was not observed with the cube. This is because the transducer was not placed at the true reattachment $RA_T$ location in the case of the cube, but at a location above the reattachment where a peak in mean pressure was observed in the preliminary static pressure measurements using the pressure scanner. There was a constraint in mounting the threaded Endevco transducers, with relatively larger diameters, at a lower height on the cubical protuberance, due to the interference with the threaded hole at the base used for mounting the protuberance on the bottom wall. Even in this case a small negative peak is observed in the correlation.

%%%%%%%%%%%%% CORRELATION:: MEAN SHOCK :: Short cylinder & Cuboid %%%%%%%%%%%%%%

%%%%%%%%%%%%%%%%%%%%%%%%%%%%%%%%%%%%%%%%%%%%%%%%%%%%%%%%%%%%%%%%%%%%%%%%%%%%

\begin{figure*}[!htb]
    \centering
    \includegraphics[width=0.95\textwidth]{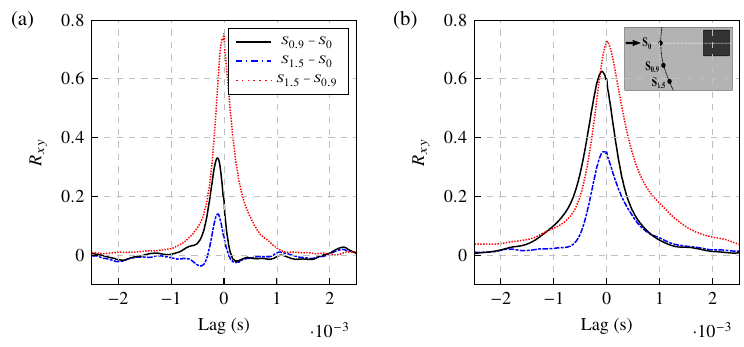}
    \caption{Mean shock correlations (a) short cylinder (b) cuboid}
    \label{figure10:correlations-mean-shock-shortcylinder-cuboid}     
\end{figure*}

To understand coherence in separation shock motion, the wall pressure signals are extracted at spanwise locations along the mean separation shock, as shown in the figure \ref{figure6:probe-locations-experiments}. Cross correlations between those locations are presented in figure \ref{figure10:correlations-mean-shock-shortcylinder-cuboid} for short cylinder and cuboid cases. The correlation between two positions is expressed as $S_{A} - S_{B}$, where $S_A$ and $S_B$ basically stand for the signals extracted (along the shock) at $z/w = A$, $z/w = B$, respectively, where $w$ is the width or side of the cube. Significant positive peaks in correlation values are observed between the signals at all probed locations for all the protuberances, suggesting that the low-frequency back-and-forth motion of the shock is coherent along the span. Concerning the correlations between centerline probe with those away from the centerline, the peak correlations remain significant but decrease with increase in the distance from centerline. It is also noticed that the correlation values vary with the extent of separation, $L_{sep}$; the correlation amplitudes are found to increase with $L_{sep}$, showing higher values for the cuboid even between locations which are far apart, and relatively lower peaks for the short cylinder. This suggests that the shock motion tends to become more coherent along the span with an increase in $L_{sep}$, a trend that is expected with reference to the asymptotic limit of the 2-D SBLI dynamics. It is significant to note that the shock foot along the sides $S_{1.5} -S_{0.9}$ is always highly correlated ( $R_{xy} \approx 0.75$), thus coherent irrespective of the protuberance, thereby of $L_{sep}$ at time lags close to zero. On the other hand, for all the locations in the intermittent region away from the centerline, for all cases, the peak value in correlation was observed at noticeable negative time lags with regard to the shock foot location at the centerline. This is indicative of a flapping motion of the shock foot, in which the upstream or downstream motion is initiated in the neighborhood of the centerline, and the corresponding movement of the portions of the shock foot away from the centerline are initiated with a time lag. Thus, despite a generally coherent to-and-fro oscillation, the shock foot experiences spanwise distortions too.

%%%%%%%%%%%%% CORRELATION:: MEAN SHOCK :: Short cylinder & Cuboid %%%%%%%%%%%%%%

%%%%%%%%%%%%%%%%%%%%%%%%%%%%%%%%%%%%%%%%%%%%%%%%%%%%%%%%%%%%%%%%%%%%%%%%%%%%

\begin{figure}
    \centering
    \includegraphics[width=0.80\linewidth]{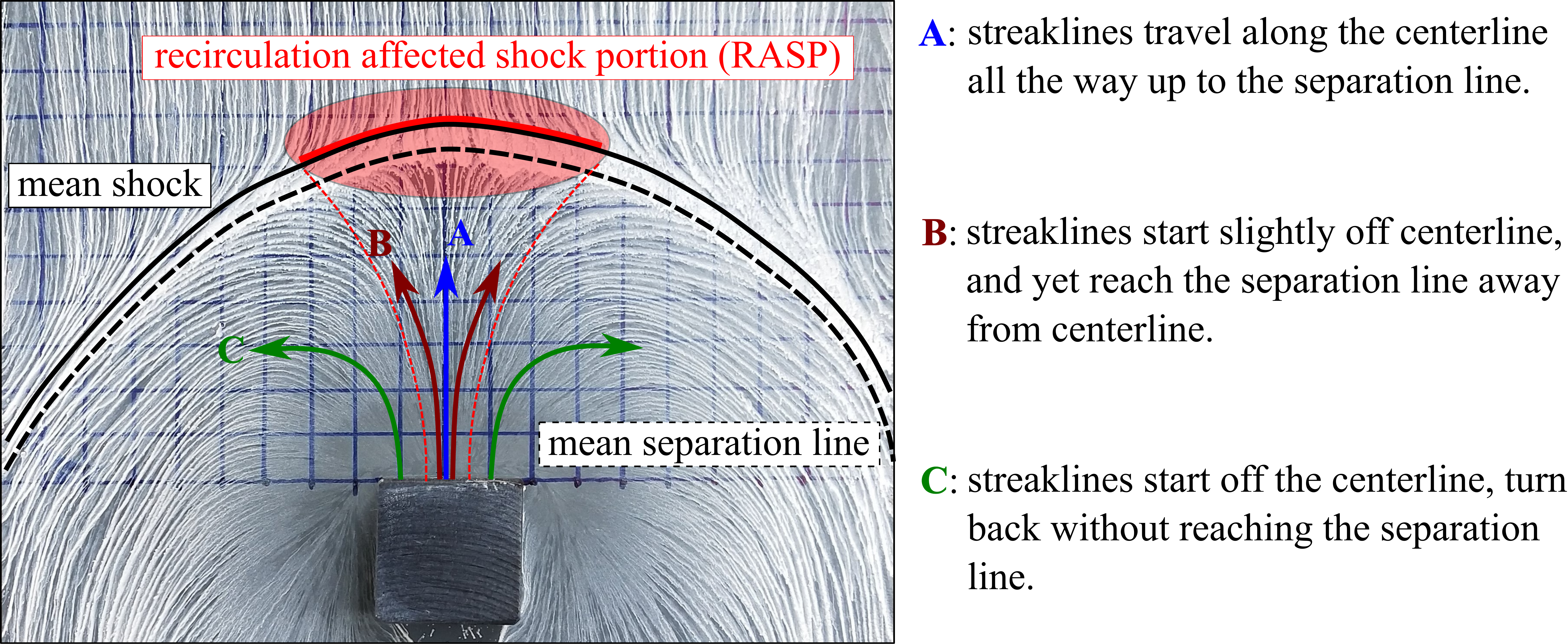}
    \caption{Streaklines affecting different regions of the interaction.}
    \label{figure11:streakline-patterns-cube-experiments}
\end{figure}

\begin{figure*}[!htb]
    \centering
    \includegraphics[width=0.95\textwidth]{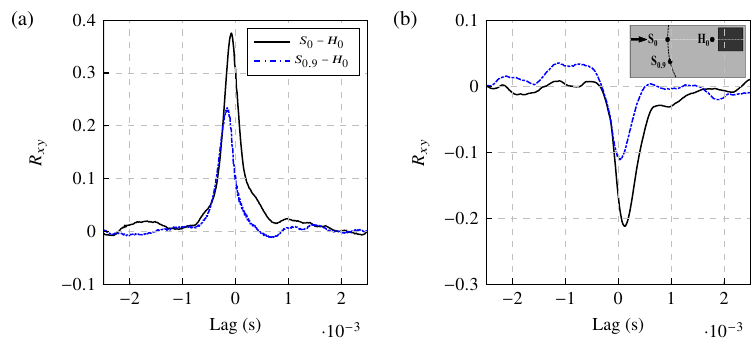}
    \caption{Mean shock and high pressure (centerline) correlations (a) short cylinder (b) cuboid}
    \label{figure12:statistical relation-MS(spanwise)-HP-experiments-shortcylinder-cuboid}    
\end{figure*}

The observation that the motion of the shock foot in the neighborhood of the centerline initiates the overall shock foot motion is further explained by the streakline patterns of the interaction, as shown in the figure \ref{figure11:streakline-patterns-cube-experiments} for the cubical protuberance. The streaklines directed upstream from the high-pressure zone near the protuberance base can be classified qualitatively into three categories: the streakline A, which goes along the centerline all the way up to the separation line; the streaklines B, which start slightly off the centerline, and yet reach the separation line at locations away from the centerline; and the streaklines C which turn back without reaching the separation line. While the line A is normal to the separation line at the point of their intersection, the angles progressively decrease between lines B and the separation line as one moves away from the centerline. Any temporal changes in mass flow along lines A and B, especially for the lines that intersect the separation line at larger angles, distort the separation line. Quantitatively, if a cut-off angle between lines B and the separation line may be fixed, it is possible to define the portion of the shock which is strongly affected by the reverse flow, referred to as `recirculation affected shock portion (RASP)' as marked in figure \ref{figure11:streakline-patterns-cube-experiments}. Indeed, significant peak correlations in surface pressures are observed between centerline location at high-pressure zone ($H_0$) and shock foot/separation locations at both centerline as well as at nearly $2\delta$ away from centerline, as shown for the cases of short cylinder and cuboid in figure \ref{figure12:statistical relation-MS(spanwise)-HP-experiments-shortcylinder-cuboid}. It can be noted that while for the short cylinder case, the peak is positive, for the case of the cuboid, it is negative. This is because, the transducer, placed at same location upstream of protuberance base, is at the region of rising pressure in the case of short cylinder whereas it is exactly at peak pressure point for the cuboid case due to larger extents of the zones with cuboid. In any case, the trends in magnitudes is the same for both cases. The observation of considerable correlations between $H_0$ and $S_{0.9}$ suggests that the fluctuations at high-pressure zone, in turn driven by fluctuations at the reattachment on protuberance face, influence directly the shock motion not just at centerline but a certain neighborhood of centerline. It is in this neighborhood that the shock motion is initiated by instantaneous fluctuations in mass flow injected from the reattachment, manifesting as pressure fluctuations near the protuberance base.

%%%%%%%%%% CORRELATION:: MEAN SHOCK-HIGH PRESSURE :: Short cylinder & Cuboid %%%%%%%%%%

%%%%%%%%%%%%%%%%%%%%%%%%%%%%%%%%%%%%%%%%%%%%%%%%%%%%%%%%%%%%%%%%%%%%%%%%%%%%

\begin{figure*}[!htb]
    \centering
    \includegraphics[width=0.95\textwidth]{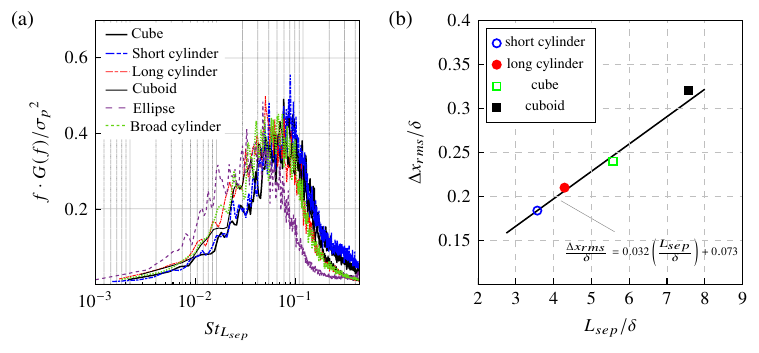}
    \caption{Shock oscillation characteristics (a) frequency scaling based on separation length applied to various protuberances. (b) root-mean-square of shock oscillations with separation extent.}
    \label{figure13:freq-scaling-shock-oscillations-experiments}    

\end{figure*}

The centerline separation length is thus a critical parameter in understanding the dynamics of the shock-induced open separation due to protuberances. Further, the $L_{sep}$ is found to be the critical parameter in determining both the length and time-scales of oscillations. When the centerline shock foot pressure spectra (PSD) for all the protuberances were plotted against the Strouhal number based on $\delta$, $St_{\delta}$, as presented in all the discussions so far, they were found to be exhibiting different ranges of dominant frequency values for different protuberances, although the dominant frequencies were of the same orders of magnitude ($St_{\delta}\sim 0.01$). However, when the spectra are plotted against Strouhal number based on centerline separation length ($L_{sep}$), $St_{L_{sep}}$, they almost collapsed into a single band with peaks around a value of  $St_{L_{sep}} \sim 0.055$ as shown in figure \ref{figure13:freq-scaling-shock-oscillations-experiments}(a). This implies that, despite different statistical features such as mean separation length and magnitudes of correlations based on differences in the extent of three-dimensional relieving, $L_{sep}$ is the key parameter in determining the time-scale of shock oscillations. Additionally, the shock oscillation amplitudes are found to vary proportionally with $L_{sep}$. It is clear from figure \ref{figure13:freq-scaling-shock-oscillations-experiments}(b) that the average shock foot oscillation amplitude, i.e, root mean square (RMS) values extracted from the scan line shock positions from schlieren images, shares a linear relation with $L_{sep}$. With these important insights from the experiments on the 3-dimensional aspects of the interaction, further details of the flow field are probed through computations, which shall be discussed subsequently.

%%%%%%%%%%%%%%%%%%%%% Centerline PSD:: Endevco %%%%%%%%%%%%%%%%%%%

 \FloatBarrier

\section{Computational Methodology}
\label{sec3:comp-methodology}

The experimental data provide valuable insights into the dynamics of the 3-dimensional separated flow field. It establishes the centerline separation length, $L_{sep}$, as a key parameter in determining shock oscillation frequencies and amplitudes. The correlations in surface pressure fluctuations allow for some important inferences on the 3-dimensional aspects of shock motion and its relation to pressure oscillations within the separated region, which, in turn, are linked to $L_{sep}$ through the RASP. However, the experiments provide a limited perspective of the flow field. A complete 3-dimensional resolution of the flow field is required to reveal the 3-dimensional nature of shock oscillation and its relation to the separated flow, especially the dynamics of the horseshoe vortex. In addition to the schlieren images (line-of-sight integration, as in experiments) providing information on shock motion in the mid-span plane, information from other planes, especially those parallel to the wall, can resolve the spanwise organization of the shock foot, which would also be beneficial for understanding the mechanism. Similarly, POD analysis based on information from the wall-parallel plane and further insights from 3-D POD would be invaluable and could be addressed using high-fidelity simulations. In the present work, an enhanced version of the DES, the 'adaptive DES' (or simply ADES), is performed for the case of the cubical protuberance to resolve 3-dimensional unsteady flow features. The ADES method offers better predictions (closer to LES) than standard DES models, due to the dynamic calculation of model coefficients, in contrast to the latter, which use constant coefficients. While the ADES model was initially developed for incompressible flows by \cite{yin2015dynamic} and \cite{yin2016adaptive}, we extended it to compressible flows by incorporating compressibility effects into the dynamic procedure and calculating the isotropic stress coefficient dynamically \citep{vayala2025adaptive}. The compressible ADES model is briefly discussed in the following section.

\subsection{Compressible ADES Model}
\label{sec3.1:ADES-model}
Unsteady, three-dimensional, Favre-filtered compressible Navier-Stokes equations are given below in Cartesian coordinates system. 

    \begin{equation}
          \frac{\partial \bar{\rho}}{\partial t} + \frac{\partial \bar{\rho} \tilde{u}_i }{\partial x_i}  = 0 
    \end{equation}
    \begin{equation}
        \frac{\partial \bar{\rho} \tilde{u}_i}{\partial t} + 
        \frac{\partial }{\partial x_j} ( \bar{\rho} \tilde{u}_i \tilde{ u}_j + \bar{p} \delta_{ij}) = 
                                                  \frac{\partial \tilde{\sigma}_{ij}}{\partial x_j} +
                                                  \frac{\partial \tau_{ij}}{\partial x_j}
    \end{equation}
    \begin{equation}
        \frac{\partial \bar{\rho}}{\partial t} + \frac{\partial }{\partial x_j} [(\bar{\rho} \tilde{E} + \bar{p}) \tilde{u}_j]  =  
                            \frac{ \partial }{\partial x_j}  [(\tilde{\sigma}_{ij} + \tau_{ij})  \tilde{u}_i] + 
                            \frac{\partial }{\partial x_j} (\tilde{q}_j + Q_{j}) 
    \end{equation}

    Here $\tilde{(.)}$ represents the Favre-filtered quantities. $\bar{\rho}, \tilde{u}_i, \bar{p}$ and $\tilde{E}$ are density, velocity vector, pressure, and total energy, respectively, and $\delta_{ij}$ is the Kronecker delta. $\tilde{\sigma}_{ij}, \tilde{q}_j$ are molecular viscous stress and heat fluxes, respectively. The turbulent stress, heat flux quantities represented by $\tau_{ij}$, $Q_{j}$, are defined as follows:
    \begin{equation}
      \tau_{ij} = \mu_{t} \left( \frac{\partial \tilde{u_i}}{\partial x_j} + \frac{\partial \tilde{u_j}}{\partial x_i} - \frac{2}{3}\delta_{ij} \frac{\partial \tilde{u_k}}{\partial x_k} \right) -\frac{2}{3}\tau_{kk} \delta_{ij}
     \end{equation}
    \begin{equation}
        Q_j = - \frac{\mu_{t} \ c_{p} }{Pr_{t}} \frac{\partial \tilde{T}}{\partial x_j}
    \end{equation}
   
   where $\mu_t$ is turbulent viscosity, $c_p$ is specific heat coefficient ($c_{p} = 1.005$ kJ/kg.K for air), and $Pr_t$ is turbulent Prandtl number ($Pr_{t} = 0.89$ for air).  $\mu_t$ is modeled using turbulence closure models, which in the present case is using the ADES model. The ADES model behaves like a conventional RANS model in the attached boundary layer regions and switches to LES based on length-scale formulations. The RANS regions are modeled using the $k-\omega$ model \citep{wilcox1988reassessment}, and the LES zones are treated with formulations equivalent to the dynamic Smagorinsky model in LES \citep{lilly1992proposed}. In addition to the main governing equations, two transport equations are solved: one for turbulent kinetic energy ($k$) and the other for the turbulent specific dissipation rate ($\omega$), providing modeled quantities for RANS and LES (sub-grid). 
    
    \begin{equation}
        \frac{\partial \bar{\rho} \tilde{k}}{\partial t} + \frac{\partial \bar{\rho} \tilde{k} \tilde{u_j}}{\partial x_j} = \tau_{ij} \frac{\partial \tilde{u_i}}{\partial x_j} - \beta^{*}\bar{\rho} \tilde{k} \tilde{\omega} + \frac{\partial }{\partial x_j} \left[ (\mu + \sigma^{*} \mu_t) \frac{\partial \tilde{k}}{\partial x_j}\right]
    \end{equation}

    \begin{equation}
         \frac{\partial \bar{\rho} \tilde{\omega}}{\partial t} + \frac{\partial \bar{\rho} \tilde{\omega} \tilde{u_j}}{\partial x_j} = \alpha \frac{\tilde{\omega}}{\tilde{k}}\tau_{ij} \frac{\partial \tilde{u_i}}{\partial x_j} - \beta \bar{\rho} \tilde{\omega}^{2} + \frac{\partial }{\partial x_j} \left[ (\mu + \sigma \mu_t) \frac{\partial \tilde{\omega}}{\partial x_j}\right]
    \end{equation}

\noindent The model coefficients are, $\alpha = \frac{5}{9}, \ \beta = \frac{3}{40}, \ \beta^* = \frac{9}{100}, \ \sigma_{k} = \frac{1}{2}, \ \sigma_{\omega} = \frac{1}{2}$. \\

When the model switches to LES approach in the separated or eddy regions, the turbulence is resolved by limiting modeled turbulent kinetic energy production through eddy viscosity formulations given below.

\begin{equation}
    \mu_t = \bar{\rho} \ell^{2}_{DDES}\tilde{\omega}
\end{equation}
\begin{equation}
  \ell_{DDES} = \ell_{RANS} - f_{d}\hspace{2mm} max( 0, \hspace{1mm} \ell_{RANS} - \ell_{LES} )
\end{equation}
\begin{equation}
  \ell_{RANS} = \frac{\sqrt{\tilde{k}}}{\tilde{\omega}}; \ \ \ \ell_{LES} = C_{DES}\Delta; \ \  C_{DES} = 0.12 
\end{equation}

\begin{equation}
    \Delta = f_{d} {V_c}^{1/3} + (1 - f_{d}) h_{max}; \ \ \ h_{max} = max (\Delta x, \Delta y, \Delta z)
\end{equation}
\\
\noindent where $\ell_{DDES}, \ell_{RANS}$, and $\ell_{LES}$ are length scales corresponding to delayed DES (DDES), RANS, and LES models, respectively. $C_{DES}$ is DES model constant. $f_{d}$ is a shielding parameter, $\Delta$ is a measure of grid, and $V_{c}$ is cell volume. \\

\noindent Isotropic stress contribution (especially for compressible flows), $\tau_{kk}$ is modeled as
\begin{equation}
      \tau_{kk} = \mu_{t_I} \tilde{\omega}
\end{equation}
where $\mu_{t_I}$ is isotropic stress viscosity, which is determined in similar fashion to $\mu_t$ with appropriate length scales. 
\begin{equation}
    \mu_{t_I} = \bar{\rho} \ell^{2}_{DDES_I}\tilde{\omega}
\end{equation}
\begin{equation}
  \ell_{DDES_I} = \ell_{RANS_I} - f_{d}\hspace{2mm} max( 0, \hspace{1mm} \ell_{RANS_I} - \ell_{LES_I} )
\end{equation}
\begin{equation}
  \ell_{RANS_I} = \frac{\sqrt{\tilde{k}}}{\tilde{\omega}} ; \hspace{5mm}  \ell_{LES_I} = C_{I}\Delta 
\end{equation}

% The RANS-LES switching is controlled using a shielding parameter ($f_d$), which prevents premature switching, which was the concern for the earlier version of DES.

%\noindent where $\Delta$ represents measure of grid in LES, $V_c$ is cell volume.

% \begin{equation}
%     f_d = 1 - \tanh([8r_d]^3)
% \end{equation}
% \begin{equation}
%     r_d = \frac{\tilde{k}/\tilde{\omega}  +  \tilde{\nu}/Re}{\kappa^2 d^2{_w} \sqrt{U_{ij} U_{ij}} }
% \end{equation}
  
% \noindent where $\tilde{\nu}$ is kinematic viscosity, $\kappa$ is von Karman constant,  $d_w$ is wall distance, $U_{ij}$ is the deformation rate tensor. 

\noindent In LES zones ($f_{d} = 1; \ \ell_{LES} < \ell_{RANS}$), the model constants, $C_{DES}$, is calculated dynamically, i.e, $C_{DES} = C_{dyn}$, using  

\begin{equation}
     C^{2}_{dyn} = max \hspace{1mm} \left(0, 0.50 \frac{\langle \ L^{*}_{ij} \ M_{ij} \ \rangle}{\langle \ M_{ij} \ M_{ij} \ \rangle} \right)
\end{equation}

\begin{equation}
    L^{\ast}_{ij} = L_{ij} - \frac{1}{3} L_{kk} \delta_{ij} \\
\end{equation}
\begin{equation}
    L_{ij}  =\widehat{\overline{\rho} \ \widetilde{u_i} \ \widetilde{u_j}}  -  \frac{\widehat{\overline{\rho} \  \widetilde{u_i}} \ \widehat{\overline{\rho} \ \widetilde{u_j} }}{\widehat{\overline{\rho}}} ; \ \ \ \ 
    M_{ij}  =  \widehat{\Delta^{2} \ \overline{\rho} \ \widetilde{\omega} \ \widetilde{S^{\ast}_{ij}}}   -   \widetilde{\Delta}^{2} \ \widehat{{\overline{\rho}}} \ \widehat{\widetilde{\omega}} \ \widehat{\widetilde{S^{\ast}_{ij}}} \\
\end{equation}
\begin{equation}
    L_{kk}  =  \widehat{\overline{\rho} \ \widetilde{u_k} \ \widetilde{u_k}}   -  \frac{\widehat{ \overline{\rho} \  \widetilde{u_k}} \ \widehat{\overline{\rho} \ \widetilde{u_k} }}{\widehat{\overline{\rho}}}; \ \ \ \ 
    S^{*}_{ij}  =  S_{ij} - \frac{1}{3}S_{kk}\delta_{ij}
\end{equation}

\noindent where $L^{\ast}_{ij}, M_{ij}, $ are resolved test-filtered stress quantities, $S_{ij}$ is the strain rate. The quantities in $\widehat{(..)}$ are test-filtered quantities and the quantities in $\langle \ \rangle$ represents the spatial-averaged quantities in the homogeneous direction if available. Test-filtering is handled using Explicit Gaussian filtering by \cite{cook2004high}, with the test-filter width to be twice the grid measure, i.e, $\widehat{\Delta} = 2\Delta$. 

\noindent Accurate calculation of $C_{dyn}$ requires a sufficient portion of the inertial sub-range resolved. On coarser meshes, as this is not feasible, a lower bound, $C_{lim}$, is set. Hence, on finer grids, the model works as dynamic DES with $C_{DES}$ attaining the $C_{dyn}$ value, while on coarser meshes, the model works as standard DDES model with $C_{DES}$ approaching $C_{DES}$ = 0.12 through $C_{lim}$ formulations given below. Therefore, the model is adaptive to the mesh, and thus called `adatpive DES (ADES)'.

\begin{equation}
    C_{DES} = max \hspace{1mm} (C_{lim}, C_{dyn})
\end{equation}
\begin{equation}
    C_{lim} = C^{0}_{DES} \left [ 1 - \tanh{\left( \alpha \exp{\left( \frac{-\beta h_{max}} {\eta} \right)} \right) } \right]
\end{equation}
   
\noindent where $C^{0}_{DES} = 0.12$, $\alpha = 25$, $\beta = 0.05$, $\eta = { \left( \frac{\nu^3}{\epsilon} \right) }^{1/4} $,  $\epsilon = 2(C^{0}_{DES} h_{max})^2 \tilde{\omega} {\mid{S}\mid}^2 + \beta^{*} \tilde{k}\tilde{\omega}$. The coefficients $\alpha$ and $\beta$ are obtained from the calibration studies of channel flows.

\noindent Isotropic stress constant, $C_I$, also computed dynamically using
\begin{equation}
      C^{2}_{I} = max \hspace{1mm} \left(0, 0.50  \frac{\langle  \ L_{kk} \ \rangle} {\langle \ \ \widehat{\Delta}^{2} \ \widehat{\overline{\rho}} \ \widehat{\widetilde{\omega}^{2}} \ - \ \widehat{ \Delta^{2} \ \overline{\rho} \   \widetilde{\omega}^{2}} \ \ \rangle} \right)
\end{equation}

%\noindent The turbulent heat flux, $Q_i$, is calculated using a constant turbulent Prandtl number, $0.89$. \\

The present compressible ADES model is implemented in a finite-difference-based in-house solver, COMP-SQUARE \citep{vadlamani2018distributed}, and validated for impinging SBLI, demonstrating its ability to capture low- and mid-frequency phenomena. The detailed model formulations (written in a curvilinear coordinate system) and validation details can be found in \cite{vayala2025adaptive}. Given the model's validation with nominal two-dimensional SBLI, the present study extends this to the three-dimensional protuberance-induced SBLI. As mentioned earlier, the computations are intended to provide a finer resolution of the flow physics rather than parametrization, and are thus carried out only for the cubical protuberance. 

\subsection{Flow configuration}
\label{sec3.1:flow-config}
\begin{figure*}[hbt!]
    \centering
    \includegraphics[width=0.85\textwidth]{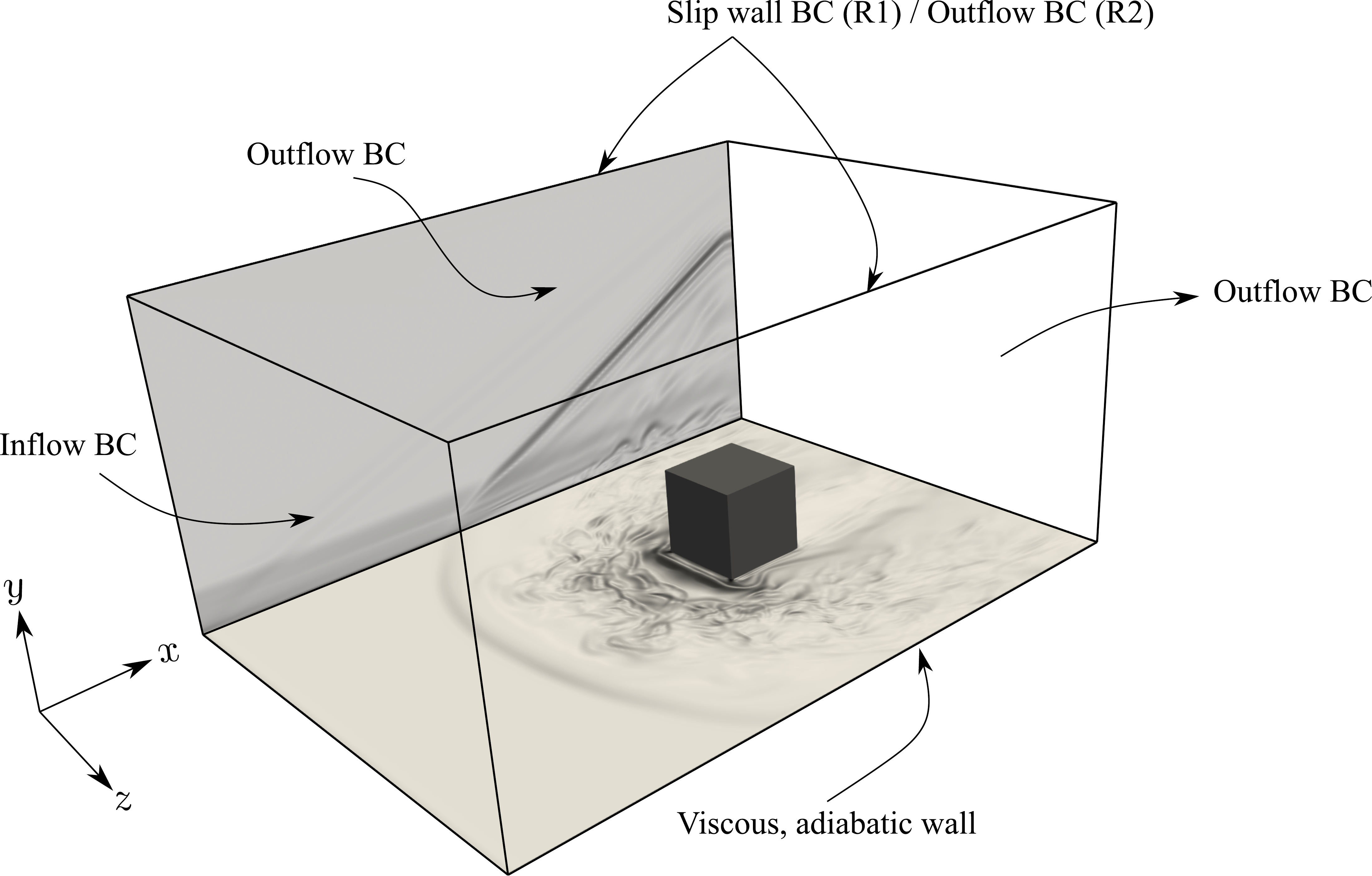}
    \caption{Computational domain with boundary conditions. Numerical schlieren on
slices ($y/\delta = 0$, $z/\delta = 0)$ are shown to illustrate the essential instantaneous flow features.}
    \label{figure1:computational domain with BCs}
\end{figure*}

 Three-dimensional shock wave boundary layer interaction due to the cubical protuberance at a freestream Mach number, $M_{\infty}$ = 2.89, is investigated numerically using adaptive detached eddy simulations (ADES). The freestream conditions provided in Table 1 are the same as those of the experiments reported in the previous section \citep{ramachandra2023study, ramachandra2023}, which are utilized as references to validate the predictions from the current study. Reynolds number based on boundary layer thickness ($\delta$) at the location of protuberance, $Re_{\delta}$, is approximately $3.3 \times 10^{5}$. A computational domain, shown in figure \ref{figure1:computational domain with BCs}, is set up with dimensions of $L_{x} \times L_{y} \times L_{z} = 17\delta \times 8\delta \times 12\delta$ in the streamwise, wall-normal, and spanwise directions, respectively. The cubical protuberance with a side of $15$ mm ($\approx 2.14\delta$) is positioned at the spanwise center ($z/\delta = 6.0$), and at a streamwise location of $x/\delta = 10$ from the inlet boundary. Concerning the boundary conditions, turbulent mean profiles (without fluctuations) obtained from the precursor RANS simulations are provided at the inlet, while at the outlet, all the quantities are extrapolated from interior points. The bottom wall is treated with non-slip, adiabatic conditions, whereas the top boundary is employed with outflow conditions. At the spanwise boundaries, two different conditions are specified, resulting in two different cases: slip wall (R1\_SPSW) and outflow (R2\_SPOF). They are also simply referred to as R1 and R2, respectively. Slip wall conditions are specified to avoid a finer resolution of side walls, thereby minimizing computational cost. Importantly, as observed in the experimental oil flow patterns, the separation bow shock becomes weak when it reaches the side walls and does not result in a strong reflected wave, and consequently, no corner separation. Therefore, the objective of having side walls is only to ensure that the confinement in experiments is replicated. Additionally, the current study explored the nature of the flow in the absence of confinement, or in other words, the effect of wall confinement by simulating the spanwise ends as outflow conditions.

%%%%%%% Table-1:: Freestream parameters %%%%%%%
 \begin{table}
    \begin{center}
    \def~{\hphantom{0}}
    \begin{tabular}{lccccccccc}
        Parameters &  $P_{0}$ & $T_{0}$ & $M_{\infty}$ & $u_{\infty}$ & $P_{\infty}$ & $\rho_{\infty}$ & $\delta$ \\ [3pt]
        Experiment \citep{ramachandra2023study} & 6 bar & 300 K & 2.87 & 612 $ms^{-1}$ & 19200 Pa & 0.598 kg$m^{-3}$ & 7 mm \\
        Computation & 6 bar & 300 K & 2.89 & 614 $ms^{-1}$ & 19193 Pa & 0.596 kg$m^{-3}$ & 7 mm\\
    \end{tabular}
    \caption{Free stream flow parameters.}
    \label{table1:freestream flow parameters}
    \end{center}
\end{table}

%%%%%%%% Computational doamin with BCs %%

%\FloatBarrier  % floats in this subsection end here
 %%%%%%%%%%%%%%%%%%%%%%%%%%%%%%%%%%%%%%%%%%%%%%%%%%

\subsection{Numerical method}
\label{Sec3.3:numerical-method}
Compressible Navier-Stokes equations are solved in conservative form in a generalized coordinate system. The compressible version of the adaptive DES model, discussed in Section \ref{sec3.1:ADES-model}, is employed to model turbulence. The governing equations are discretized spatially using explicit fourth-order, while the temporal evolution of the flow is achieved through the Runge-Kutta fourth-order (RK4). The cube zone in the computational domain is handled using the boundary data immersion method (BDIM) \citep{schlanderer2017boundary}; boundary conditions on the cube region are obtained via a meta equation that relates the Navier-Stokes equations in the fluid region to the governing equations in the solid region. An implicit Pade-type filter with low-dissipation characteristics is applied after each timestep to minimize instabilities that arise during flow evolution. An eighth-order filtering (F8) with a filtering coefficient, $\alpha_{f}$, of 0.45 is provided, except in the neighborhood of domain boundaries, where the filtering orders are reduced towards the boundaries, leaving boundary points unfiltered. Turbulent quantities ($k, \omega$) are filtered using second-order (F2) filtering, owing to their stiffness towards the higher-order schemes. A shock-capturing method based on the adaptive filtering approach with a fifth-order WENO sensor is used to capture shock waves and other discontinuities \citep{visbal2005shock}.
The computational domain as mentioned earlier, $L_{x} \times L_{y} \times L_{z} = 17\delta \times 8\delta \times 12\delta$, is discretized with $N_x \times N_y \times N_z$ = $255 \times 175 \times 337$ grids in streamwise, wall-normal, and spanwise directions, respectively. In the streamwise direction, the grid is distributed non-uniformly, with finer resolution in the interaction region. Streamwise resolution of $\Delta x^+_{min} \approx 45 $ occurs near the cube, while $\Delta x^+_{max} \approx 200 $ exists at the outflow boundary due to gradual stretching. Near-wall resolution of $\Delta y^{+} \approx 1.20$ is ensured, and the grids are further distributed using hyperbolic stretching within the boundary layer, transitioning to uniform stretching at the boundary layer edge and continuing to the top boundary. A uniform meshing is applied in the spanwise directions with $\Delta z^{+} \approx 45$. The grid resolutions used in the current study meet the requirements of LES, as mentioned in literature \citep{georgiadis2010large}. The simulations are carried out with a time resolution, $\Delta t u_{\infty}/\delta = 1 \times 10^{-3}$, which corresponds to Courant-Friedrichs-Lewy (CFL) number of 0.8. After flushing out the initial transients for six flow-through times (FTT), the simulations are continued for an additional 17 FTTs to achieve statistical steady state, after which the data are recorded for post-processing. Data probes are employed (based on time-averaged flow field) in the critical flow regions, including the pressure measurement locations on the wall in the experiments, to facilitate the comparison. The signals are extracted at 1 MHz for another 120 FTTs, yielding an ensemble sample size of 24000.  

\section{Observations from computations}
\label{Sec4:comp-observations}

%%%%%%%%%%%%%%%%%%%%%%%%%%%%%%%%%%%%%%%%%%%%%%%%%%%%%%%%%%%%%%%%%%%%%%%%%%%%%%%%%%%%%%%%%%%%%%%%%%%%%%%%%%%%
\subsection{Mean flow organization}
\label{Sec4.1:mean-flow-org}
\begin{figure*}[!htb]
    \centering  
    
    \begin{tikzpicture}
        \node[anchor=north west] at (0,0) (plotA)
        {
        \includegraphics[width=0.70\textwidth]{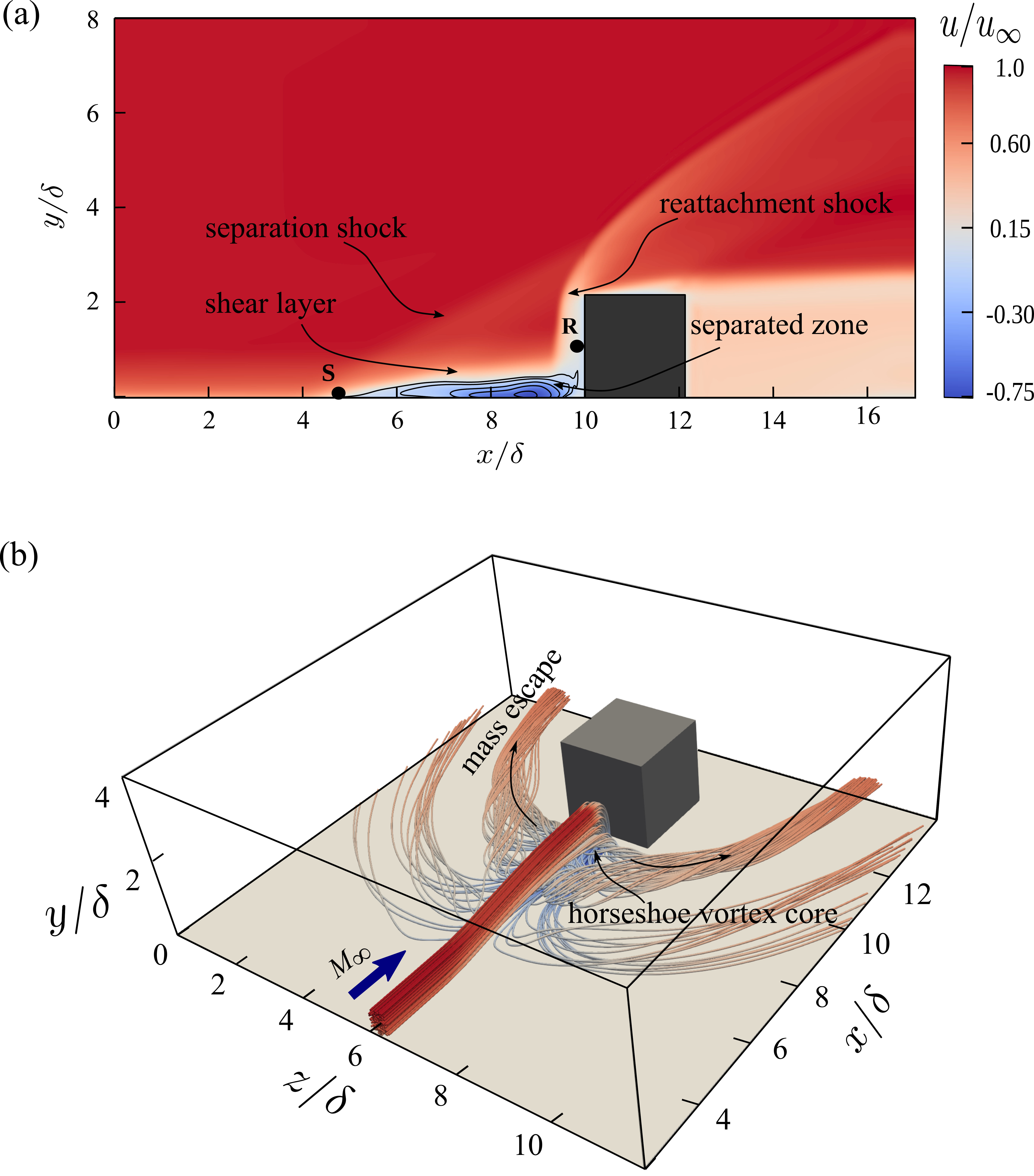}
        };
        %\node at (plotA.north west) {(a)};
    \end{tikzpicture}
    \caption{Mean flow patterns of (a) streamwise velocity (b) streamlines originating at mid-span from the upstream (inflow) boundary.}
    \label{figure2:mean flow-velocity-streamlines}    
    
\end{figure*}

The mean (averaged) flow field is firstly explained by means of velocity field shown in figure \ref{figure2:mean flow-velocity-streamlines}(a) extracted in the x-y plane at centerline span. From the averaged velocity field, it is evident that the current computations capture the critical flow features such as shocks, separated zones, etc., which are qualitatively comparable to the mean flow features seen in the present experiments as well as those reported in literature \citep{hung1985simulation, lakshmanan1994investigation}. The separation point S and the reattachment point R are also marked on the figure, and the mean shock foot on the wall can be seen upstream of S. Point R is roughly at the mid-point of the protuberance face for the present case. Figure \ref{figure2:mean flow-velocity-streamlines}(b) shows the spatial evolution of a few streamlines in the mean flow field, starting from the inlet at mid-span, illustrating the open separation in which, the mass entering the separated region escape sideways (spanwise) through the horseshoe vortex core.

%% 3-D STBLI :: Mid-span velocity and schlieren %%

%% 3-D STBLI :: Distinct pressure zones of STBLI %%
\begin{figure*}[hbt!]
    \centering
    \includegraphics[width=0.55\textwidth]{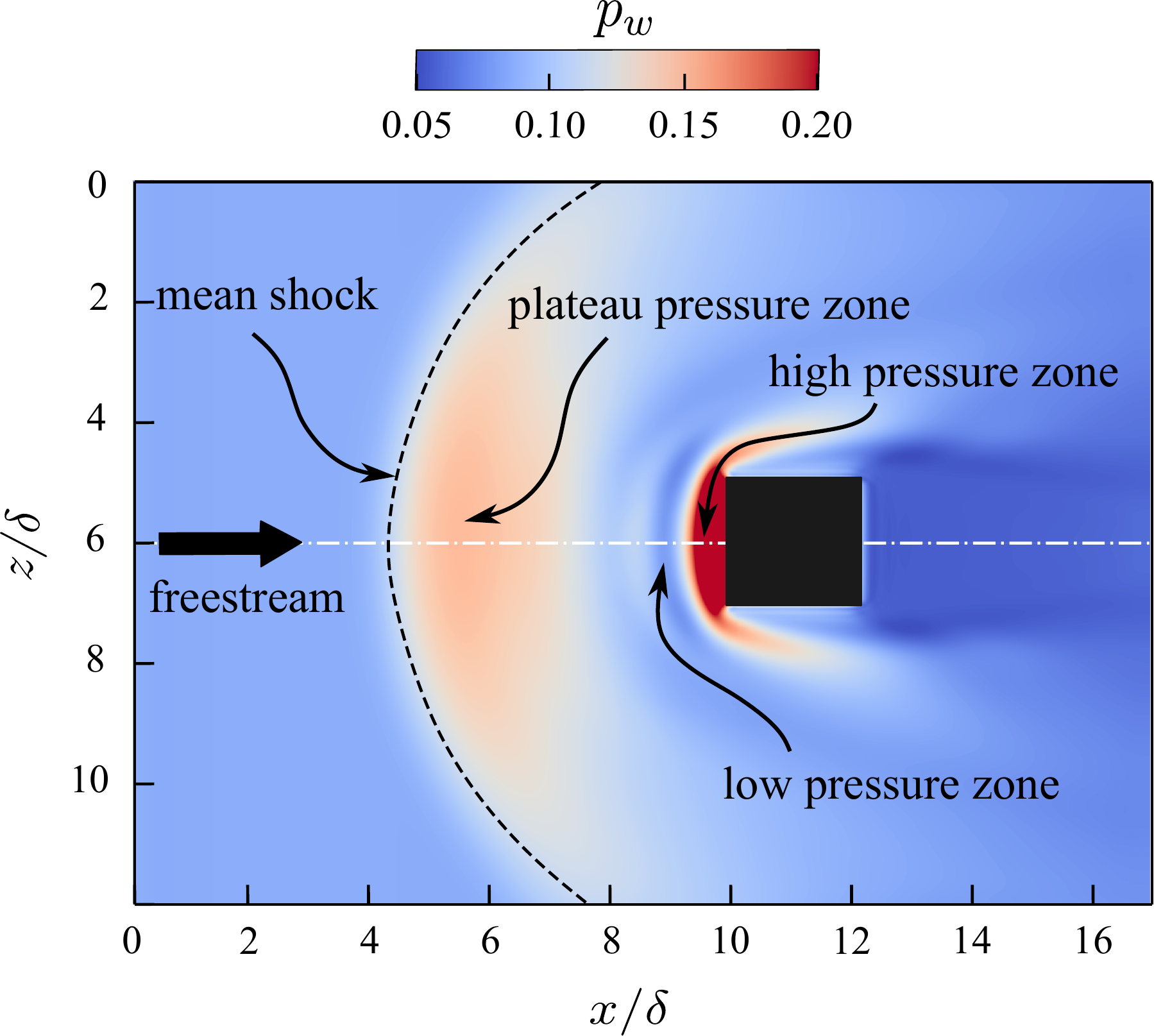}
    \caption{Contours of surface pressure. Distinct pressure zones of the interaction are marked.}
    \label{figure3:pressure zones}
\end{figure*}

Further, a spanwise view of the interaction is illustrated through the mean surface (wall) pressure field in figure \ref{figure3:pressure zones}. The qualitative similarities with the experiments are apparent with the wall pressure distribution too. As discussed earlier, the interaction is characterized by distinct pressure zones such as freestream, plateau pressure, low-pressure, and high-pressure zones. It must be added that, despite exhibiting spatial asymmetry about the span centerline for most instances, all the time averaged flow features such as the mean shock foot and the wall pressure distribution are symmetric about the span centerline.

%%%%%%%%%%%%%%%%%%%%%%%%%%%%%%%%%%%%%%%%%%%%%%%%%%%%%%%%%%%%%%%%%%%%%%%%%%%%%%%%%%%%%%%%%%%%%%%%%%%%%%%%%%%%

\subsection{Instantaneous flow organization}
\label{Sec4.2:instantaneous-flow-org}
\begin{figure*}[hbt!]
\centering
\includegraphics[width=0.65\textwidth]{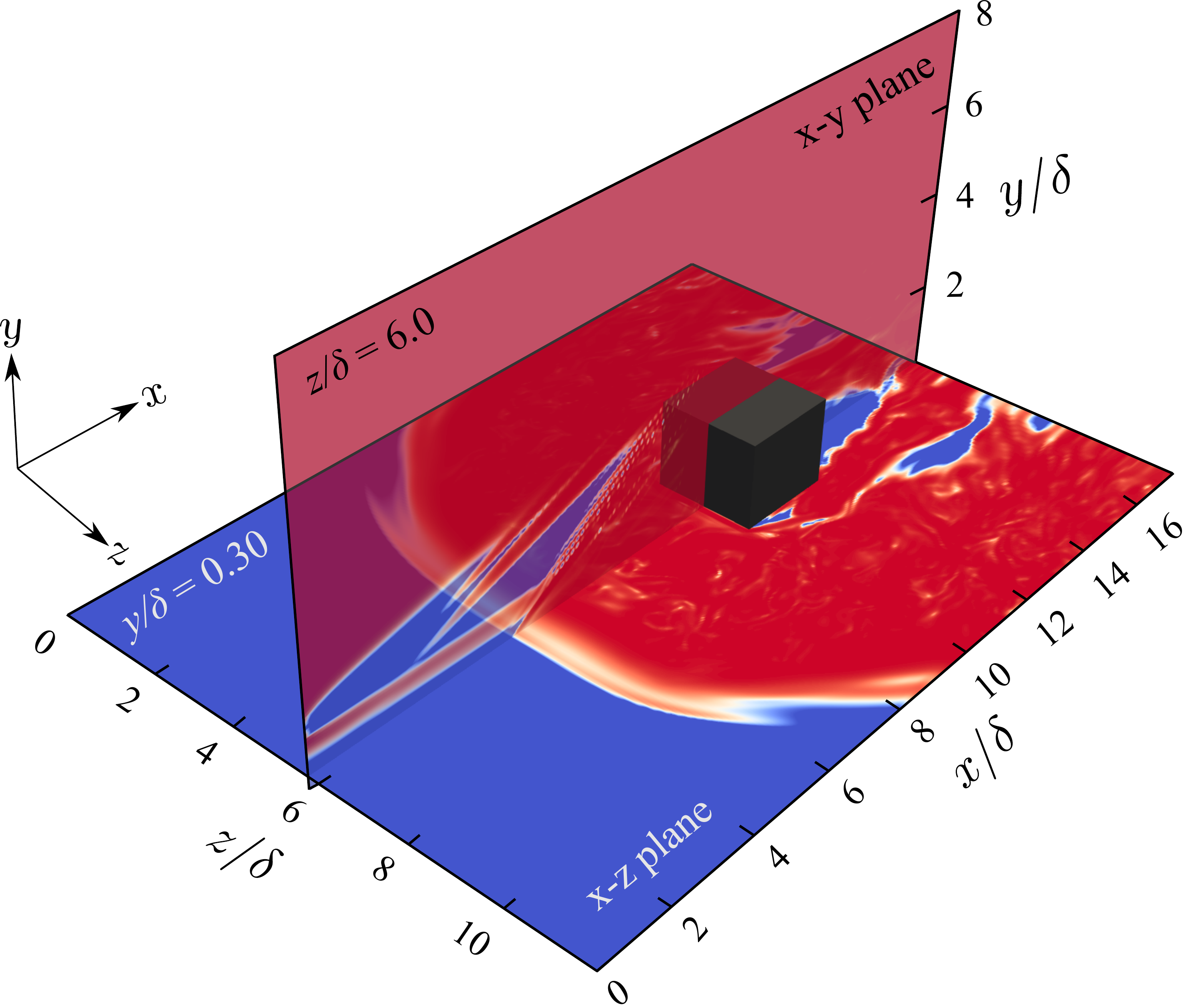}
\caption{Fields of $L_{R}$ representing RANS ($L_{R} = 0, \,\ blue$) and LES ($L_{R} = 1, \,\ red$) zones extracted along $y/\delta = 0.3$ and $z/\delta = 6.0$}
\label{figure4:LR contour}
\end{figure*}

The extent to which the instantaneous flow is resolved or modeled depends on the mode (RANS/LES) of the ADES model that operates in the flow domain. Regions where the model operates in RANS mode result in averaged flow, while the zones with LES mode can resolve the motion of various scales including turbulence. Figure \ref{figure4:LR contour} shows an instantaneous field of a parameter, $L_R$, depicting RANS ($L_R = 0$) and LES ($L_R = 1$) zones. Two different planar slices (x-z, x-y), one at $y/\delta = 0.30$ and another at $z/\delta = 6.0$, are chosen to demonstrate the RANS/LES switching of the ADES model. It is observed that in the upstream region ($x/\delta < 4.30$) ahead of the interaction, the boundary layer is primarily modeled using RANS equations, although it includes bands of LES regions. Downstream of the interaction onset or separation shock, the model switches to LES mode and resolves turbulent eddies. The LES mode remains active in almost all downstream regions, except in the wake region behind it.

\begin{figure*}[hbt!]
\centering
\includegraphics[width=0.70\textwidth]{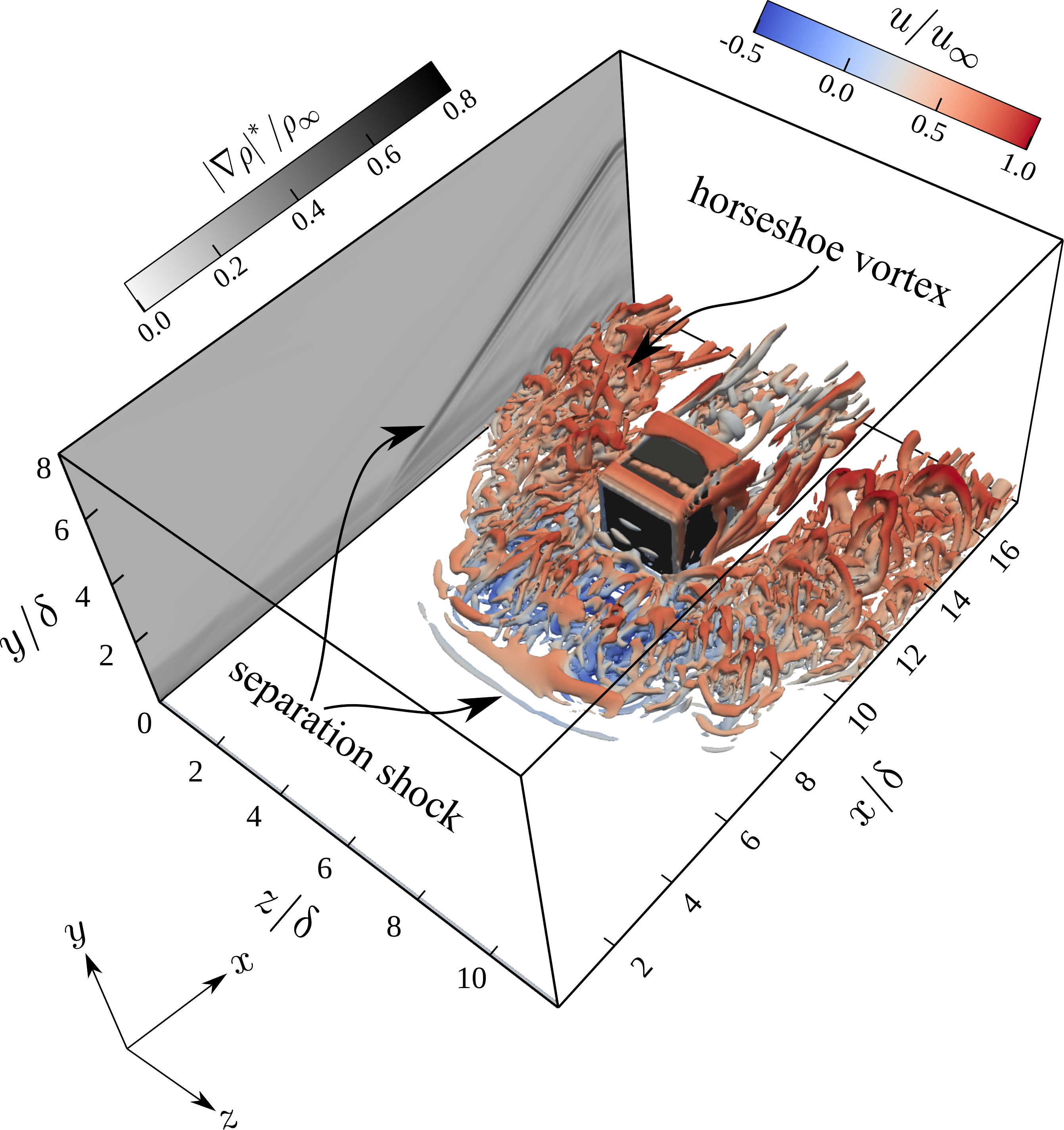}
\caption{Iso-surfaces of Q-criterion (Q = 0.1) coloured with streamwise velocity. A numerical schlieren on a slice ($z/\delta = 0$) is also included, where $|\nabla\rho|^{*} = 0.80 \times e^{-15 |\nabla \rho|/ |\nabla \rho|_{max}} $.}
\label{figure5:Qcirterion-turbulnet structures}
\end{figure*}

The coherent structures resolved by the present model are visualized using the Q-criterion, and one such instantaneous realization is shown in figure \ref{figure5:Qcirterion-turbulnet structures}. Numerical schlieren at one of the spanwise boundaries ($z/\delta = 0$) is also shown in the figure. As the inflow is provided with turbulent mean profiles, no turbulent scale generation is observed until the separation region commences. Following the separation, the model activates the LES mode, thereby resolving the different flow scales in almost the entire domain downstream of separation, including the interaction zone and further downstream of the cube face. The separation shock from the numerical schlieren slice at $z/\delta = 0$ also indicates the onset of the LES mode as the interaction starts. 

%\FloatBarrier  % floats in this subsection end here
%%%%%%%%%%%%%%%%%%%%%%%%%%%%%%%%%%%%%%%%%%%%%%%%%%%%%%%%%%%%%%%%%%%%%%%%%%%%%%%%%%%%%%%%%%%%%%%%%%%%%%%%%%%%

\subsection{Validation of computational results with experiments}
\label{Sec4.3:validation-of-computations}
%% write a small paragraph about overview of the results
To assess the efficacy of the current numerical methodology, the computational results are validated against experiments under similar flow conditions, as listed in Table \ref{table1:freestream flow parameters}. The results from both R1 and R2 computational cases are compared with the experiments. In addition to mean flow features, the unsteady characteristics of the separation shock are also compared to corroborate the agreement between computations and experiments. 
 
\subsubsection{Mean flow features}
\label{Sec4.3.1:mean-flow-features}
\begin{figure*}[hbt!]
    \centering
    \includegraphics[width=0.95\textwidth]{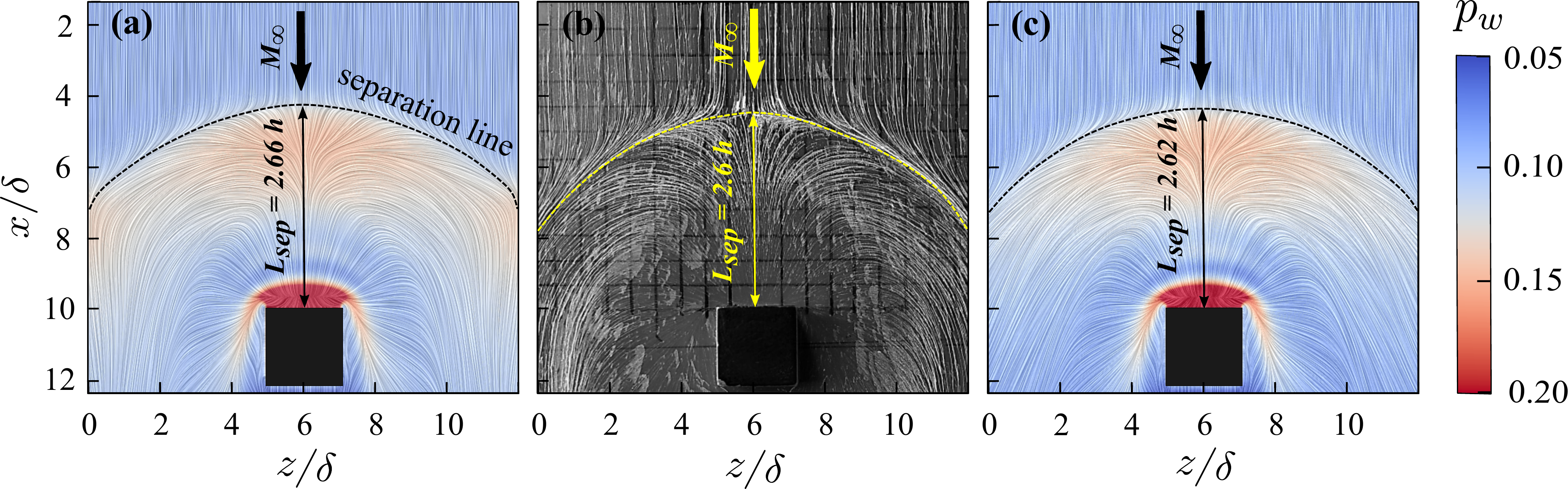}
    \caption{Topology of the mean flow (a) surface streaklines (R1) (b) oil flow visualization (experiments) (c) surface streaklines (R2)}
    \label{figure6:streaklines comparison}
\end{figure*}

The topology of the mean flow is examined relative to experiments using streaklines on the wall. For this purpose, surface streaklines from the numerical results are constructed and compared with oil flow visualization from the experiments. From the figure \ref{figure6:streaklines comparison}, it is clear that the flow patterns from the simulations are consistent with experimental oil flow patterns. However, slight variations in the directions of streak lines near the spanwise edges are observed in the simulations with the spanwise outflow (R2) case. Such minimal deviations are attributed to the difference in the conditions applied at the spanwise boundaries. Nonetheless, such disparities are found to have a negligible effect on the interaction characteristics, as discussed in the later sections. From the comparison of separation extent with experiments, the near-accurate prediction of mean separation length along the spanwise centerline ($L_{sep}=2.66h$, where \textit{h} is the height of the cubical protuberance) is apparent from the numerical surface streak lines (for R1) in comparison with oil flow visualization ($L_{sep}=2.6h$) of experiments. Interestingly, the simulations with spanwise outflow conditions results marginally smaller separation ($L_{sep}=2.62h$), which can be attributed to the three-dimensional relieving effects.

% Streaklines :: Experiments and Computations %%

\FloatBarrier  % floats in this subsection end here

\begin{figure*}[!htb]
    \centering
    \includegraphics[width=0.75\textwidth]{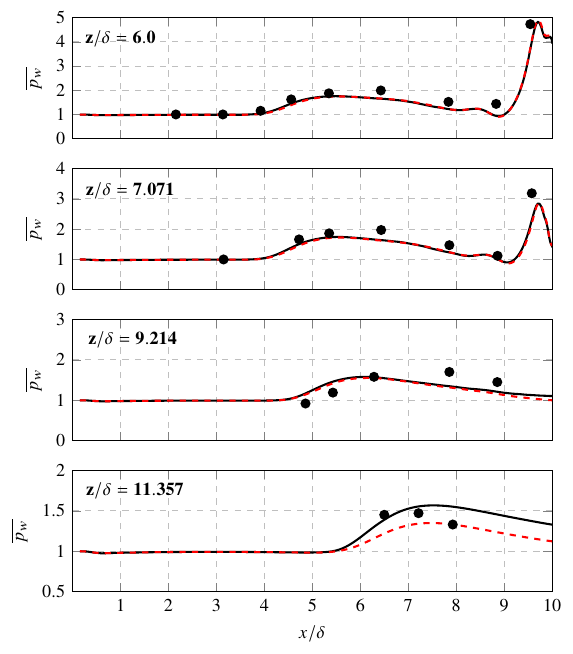}
    \caption{ Comparison of mean surface pressure distributions at various span locations: filled circle - experiments; solid line - computations (R1); dotted line - computations (R2).}
    \label{figure7: wall pressure comparision}
\end{figure*}

Figure \ref{figure7: wall pressure comparision} compares the streamwise distribution of mean wall pressure extracted at four different spanwise sections. It is clear that the numerical outcomes (R1 and R2) are very much consistent with experimental data along the centerline and vicinity regions, whereas in distant regions, some discrepancies are observed. In most spanwise locations, the pressure distribution is nearly the same for both R1 and R2 cases. However, at locations close to the spanwise boundary, the pressures were lower with the outflow condition (R2) when compared to that with the slip wall confinement (R1). 

%%% Mean Surface Pressure Distribution:: Experiment and Computation at z = 0 (centerline) %%%

%%%%%%%%%%%%%%%%%%%%%%%%%%%%%%%%%%%%%%%%%%%%%%%%%%%%%%%%%%%%%%%%%%%%%%%%%%%%%%%%%%%%%%%%%%%%%%%%%%%%%%%%%%%%%%%%%

\subsubsection{Separation shock oscillations}
\label{Sec4.3.2:sep-shock-oscillations}
\begin{figure*}[hbt!]
    \centering
    \includegraphics[width=0.75\textwidth]{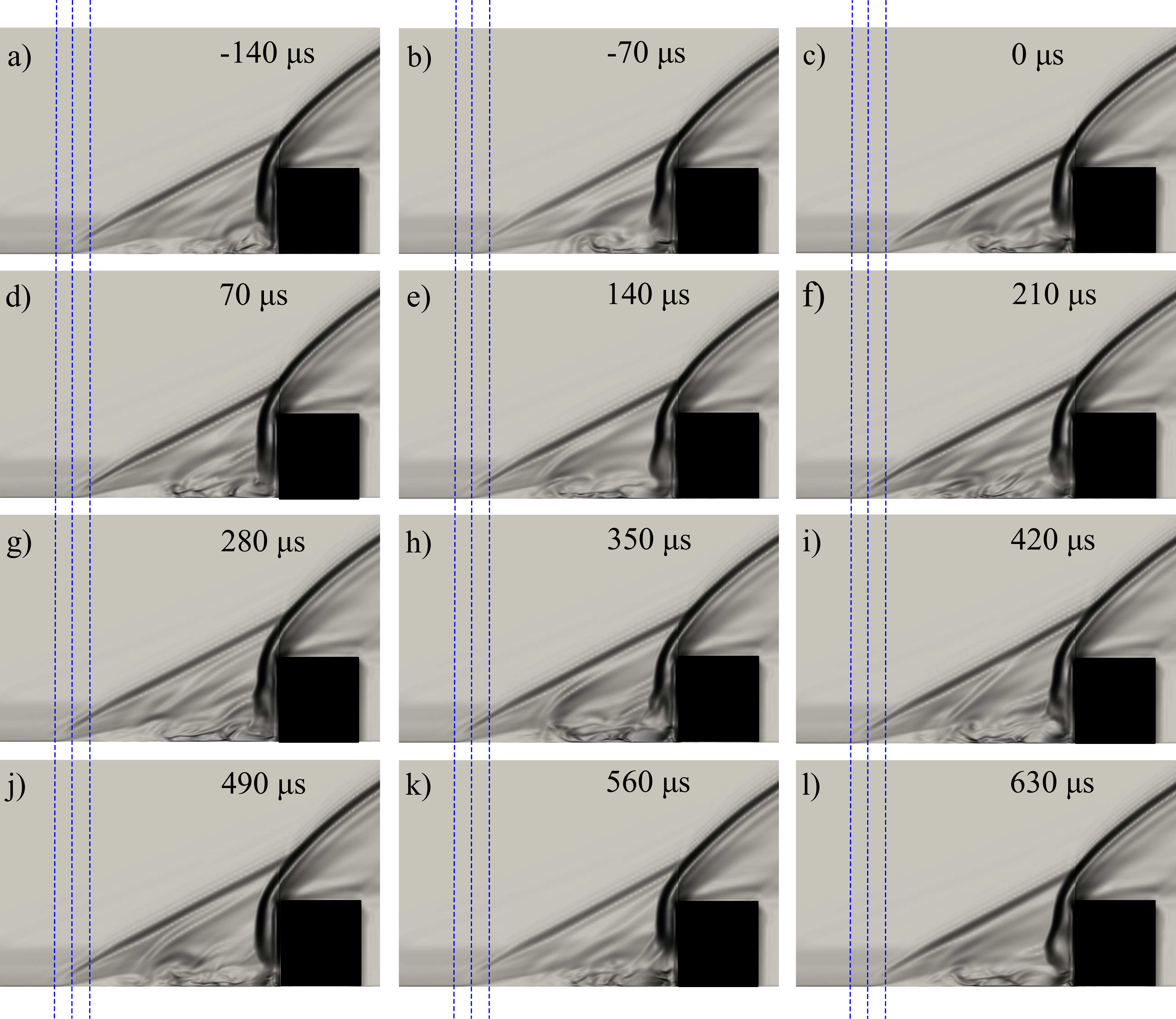}
    \caption{Instantaneous numerical schlieren snapshots illustrating the temporal excursions of the separation shock along the centerline span. Three dotted lines denote the upstream-most, approximate mean, and downstream-most positions of the shock foot, respectively.}
    \label{figure8:temporal evolution of centerline shock}
\end{figure*}

Instantaneous schlieren snapshots from simulation R1, which constitutes one (low-frequency) oscillation cycle, are presented in figure \ref{figure8:temporal evolution of centerline shock}. The flow instances at every $70\,\upmu \text{s}$ are provided along with two instances (at $t = -140\,\upmu\text{s}, -70\,\upmu\text{s}$) corresponding to the culmination phase of the previous cycle. At the initial stage, $t = 0\,\upmu\text{s}$, the shock foot is at the downstream-most position (closest to the protuberance) and starts moving upstream in subsequent instances ($t = 70\,\upmu\text{s}, 140\,\upmu\text{s} \ \& \ 210\,\upmu\text{s}$), reaching the upstream-most position at $280\,\upmu\text{s}$, completing approximately half a cycle. Following this, the shock undergoes rearward/downstream motion, eventually reaching the downstream-most position, roughly where it initiated its upstream motion, thereby completing the cycle. Three dotted lines denote the upstream-most, approximate mean, and downstream-most positions of the separation shock foot, respectively. Corresponding to the time taken for completing this cycle, the non-dimensional frequency (Strouhal number based on boundary layer thickness ($\delta$), $St_{\delta}$) is calculated to be 0.018, thereby confirming the low-frequency nature of oscillations ($St_{\delta} \sim 0.01$).

\begin{figure*}[hbt!]
    \centering
    \includegraphics[width=0.75\textwidth]{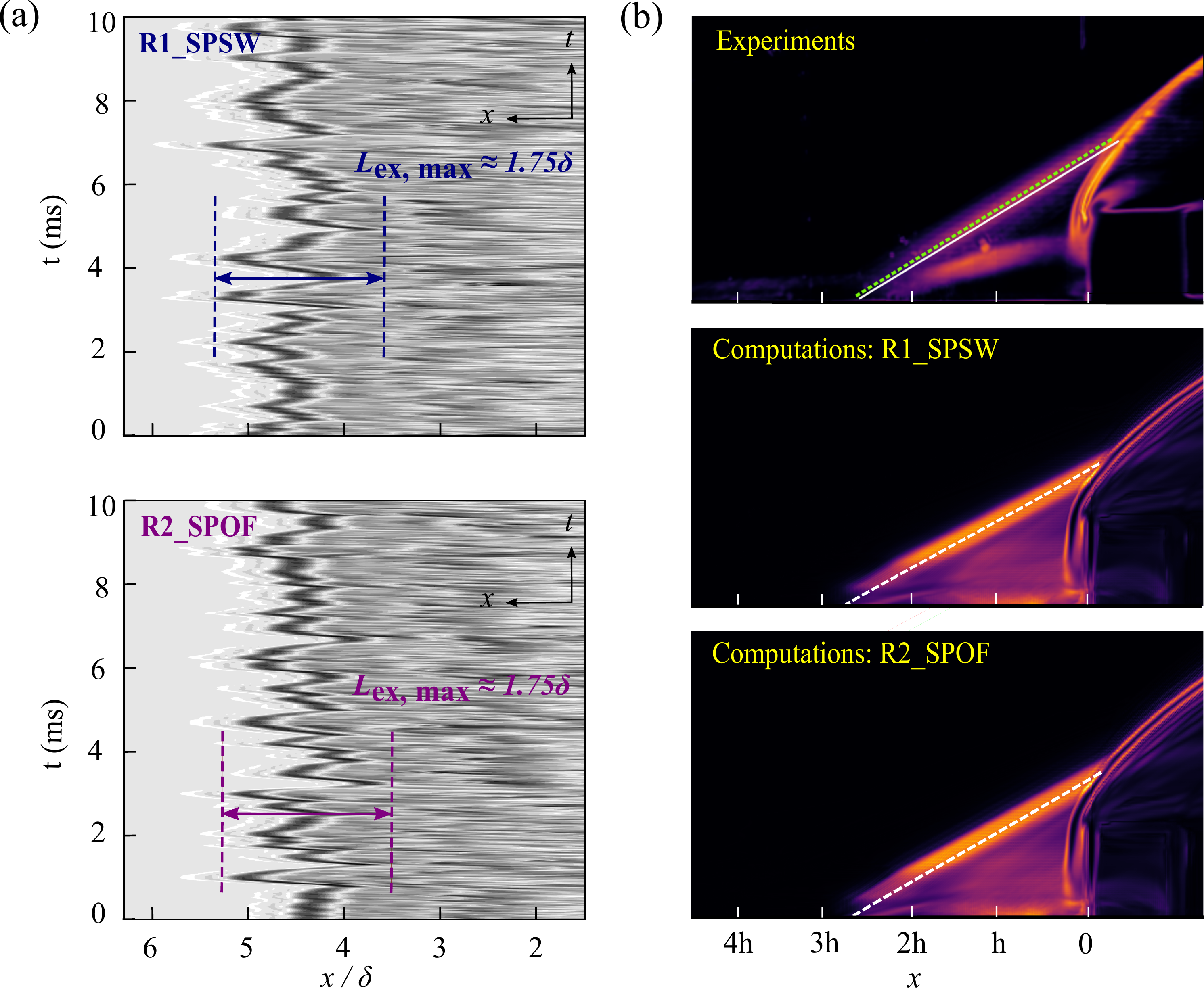}
    \caption{Centerline shock characteristics (a) space-time (x-t) variation (b) standard deviation of pixel intensities in numerical schlieren.}
    \label{figure9:centerline_x-t_standard-dev}
\end{figure*}

 To quantify the extent of shock oscillations, space-time (x-t) variation of the density gradient (schlieren) snapshots is plotted along the centerline span. Pixel intensities are extracted (for each snapshot) along a scanline at a height of y = $0.92\delta$ from the base plate, similar to experiments. Schlieren snapshots are collected at a rate of $100$ kHz for the duration of $2100\delta/u_{\infty}$ (24 ms), resulting in $2400$ data samples. Figure \ref{figure9:centerline_x-t_standard-dev}(a) illustrates that the shock motions from computations (R1 and R2) are qualitatively similar to each other, with a similar magnitude of shock excursions. These shock oscillations are qualitatively similar to those observed in experiments. However, the maximum excursion of shock, $L_{ex, max}=1.75\delta$, is under-predicted by computations, when compared with experiments where it was observed to be around $2.50\delta$. However, the standard deviation of shock foot position is found to be $\Delta x_{rms}/\delta \approx 0.27$ with the computations, which matches with the experimentally determined value (see figure \ref{figure13:freq-scaling-shock-oscillations-experiments}(b)). The noted maximum excursions are based on shock foot positions at certain instances, and a longer observation time in the computations could possibly result in larger maxima in shock excursions. The standard deviations of pixel intensities, showing the extent of shock oscillations, from both experiments and computations are also compared and found to be in good agreement, as shown in the figure \ref{figure9:centerline_x-t_standard-dev}(b). It is clear that the type of boundary condition at the span edges has a negligible effect on the statistical aspects of the interaction along the spanwise centerline.

%%% x-t Plot :: Experiments and Computations %%%

%\FloatBarrier  % floats in this subsection end here

%%%%%%%%%%%%%%%%%%%%%%%%%%%%%%%%%%%%%%%%%%%%%%%%%%%%%%%%%%%%%%%%%%%%%%%%%%%%%%%%%%%%%%%%%%%%%%%%%%%%%%%%%%%%%%%%%%
%%%%%%%%%%%%%%%%%%%%%%%%%%%%%%%%%%%%%%%%%%%%%%%%%%%%%%%%%%%%%%%%%%%%%%%%%%%%%%%%%%%%%%%%%%%%%%%%%%%%%%%%%%%%%%%%%%

\subsection{Spectral analysis}
\label{Sec4.4:spectral-analysis}
\begin{figure*}[hbt!]
    \centering
    \includegraphics[width=0.95\textwidth]{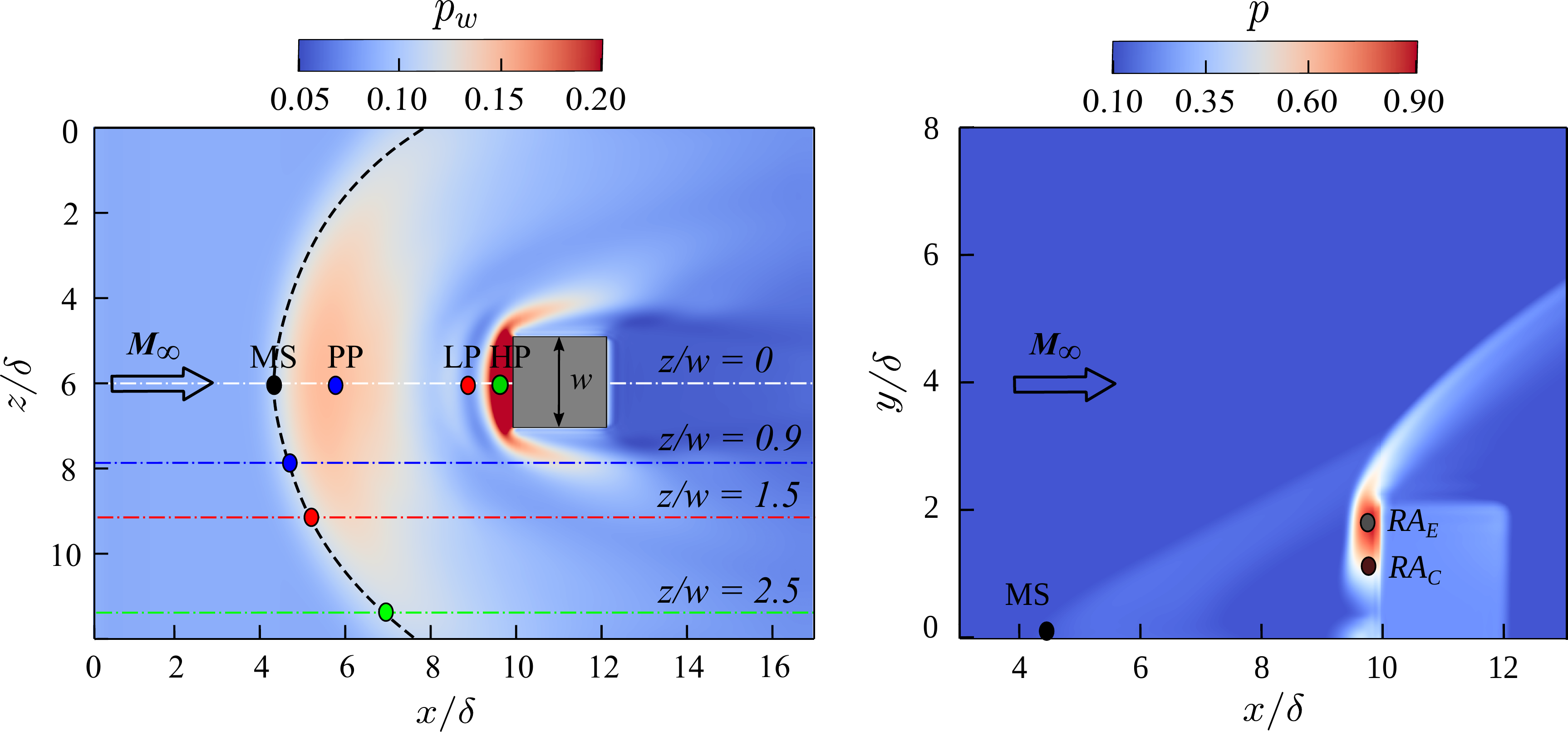}
    \caption{Distribution of data probes in the interaction zone ( MS - mean shock, PP - plateau pressure, LP - low pressure, HP - high pressure, $RA_{E}$ - reattachment (experiments), $RA_{C}$ - reattachment (computations) )}
    \label{figure10:data-probes-interaction-zone}
\end{figure*}

The unsteady aspects of the interaction are characterized by performing spectral analysis of signals probed at various locations (including experimentally measured locations) spanning the interaction zone. A set of pressure signals is extracted at the spanwise centerline to characterize spectra as well as the correlations of pressure fluctuations in the intermittent zone (in the vicinity of the mean shock foot location) with those in other zones of the interaction. Signals are also extracted over different spanwise locations along the mean shock foot, as with the experiments, to establish the existence of spanwise coherence in shock motion. Additionally, data probes are deployed at the reattachment location along the centerline span ($z/\delta = 6.0$). Figure \ref{figure10:data-probes-interaction-zone} shows the selected probe locations used in computations. The locations on the bottom wall are the same as those in the experiments with cubical protuberance \citep{ramachandra2023study}. It is worth noting that the reattachment location considered in the present experiments, $RA_{E}$, differs from the computational one, $RA_{C}$, and is located at a relatively larger height. The experimental probe location was chosen as the peak pressure point (at $y/\delta = 1.72$) based on preliminary mean surface pressure measurements, and due to the constraint in mounting the threaded Endevco transducer; however, from the computations (which enable precise characterization of reattachment point using the skin friction value), the reattachment was found to be at $y/\delta = 1.14$, considerably below the peak pressure location. The signals are recorded during the simulation at a rate of 1 MHz, yielding a total of 24000 samples, for a computational time of $2100 \delta/u_{\infty}$. Power spectral density is computed using the \textit{Welch} algorithm with a \textit{Hamming} window of size 24000 and 50\% overlap. PSD and coherence values are plotted against the non-dimensional frequency, i.e, Strouhal number ($St_{\delta}$), based on $\delta$. The spectra and correlations at locations similar to those in experiments are chosen to facilitate comparison and reinforce the findings.    

%%% x-t Plot :: Experiments and Computations %%%

\begin{figure*}[!htb]
    \centering
    \includegraphics[width=0.95\textwidth]{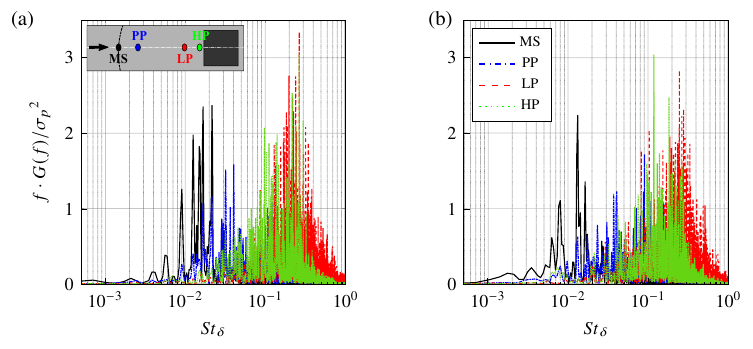}
    \caption{Power spectral densities of probes along the centerline span (a) R1\_SPSW (b) R2\_SPOF}
    \label{figure11:centerline-probes-psd}    

\end{figure*}

%%%%%%%%%%%%%%%%%%%%%%%%%%%%%%%%%%% CENTERLINE SHOCK CHARACTERISTICS %%%%%%%%%%%%%%%%%%%%%%%%%%%%%%%%%%%%%%%%%%%%
% \subsubsection{Centerline shock foot spectra and correlations}
% \label{Sec4D-1:centerline-spectra}
Regarding the centerline ($z/\delta = 6.0$) shock characteristics, the pressure signals are extracted at the mean shock foot (MS), plateau pressure (PP), low-pressure (LP), and high-pressure (HP) zones, as indicated in the figure \ref{figure11:centerline-probes-psd}, which shows the PSD of signals at these locations on the centerline. The spectra, for both R1 and R2, at the respective points, are qualitatively similar to the experimental observations, with the dominance of low frequencies, i.e., $St_{\delta} \sim 0.01$, at the mean shock location, and mid-frequencies, i.e., $St_{\delta} \sim 0.1$, dominating the other regions. There is no appreciable energy in the mid- or high-frequency ($St_{\delta} \sim 1$) ranges at the mean shock location, but at other regions, noticeable energy is apparent in the low-frequency ranges. While the PSDs are qualitatively similar for both slip wall and outflow cases, some differences in amplitude at some distinct frequencies can be observed at the mean shock foot location. Nonetheless, this does not affect the trends in correlations or coherence of pressure fluctuations between different locations. Hence, we present and discuss only the R1\_SPSW results subsequently, unless any considerable difference observed between the cases in some aspect needs to be mentioned.

%%%%%%%%%%%% Different Pressure Zones & PSD %%%%%%%%%%%%
\begin{figure*}[!htb]
    \centering
    \includegraphics[width=0.95\textwidth]{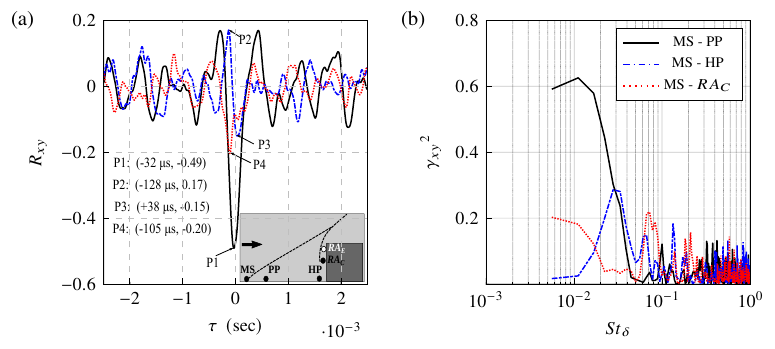}
    \caption{Statistical relations between mean shock and other locations (a) cross correlation. (b) coherence.}
    \label{figure12:centerline-statistical relation-MS-PP-HP-RR}    
    
\end{figure*}

%%%%%%%%%%%%%%%%%%%%%%%%%%%%%%%%%%%%%%%%%%%%%%%%%%%%%%%%%%%%%%%%
In correlations and coherences too, the computations capture the trends observed in the experiments (see figures \ref{figure8:centerline-statistical relation-MS-LP-experiments} and \ref{figure9:correlation-sep-reat-experiments}). Figure \ref{figure12:centerline-statistical relation-MS-PP-HP-RR} shows the cross-correlations and coherences between the signal at the mean shock foot location (MS) and other regions (PP, HP, and $RA_{C}$). A large negative correlation peak at a negative time lag (denoted as P1) is observed in the correlation between signals at MS and PP as well as LP (MS - LP correlation is not shown in the figure), which is consistent with the experimental observations at corresponding locations. High coherences are observed for lower frequencies ($St_{\delta} \sim 0.01$), suggesting a high degree of linear relationship between the low-frequency shock oscillations and the pressure fluctuations at PP and LP. At other locations too significant coherences with the signal at MS are observed in the low-frequency ranges. The following two deviations in the correlations are noteworthy. The MS-HP correlation has a considerable positive peak (denoted as P2) at a negative time lag of about 130 $\upmu$s, and a comparable negative peak (denoted as P3) at a smaller positive time lag. In the experiments (reported in \cite{ramachandra2023study}) though, for the same MS and HP locations, the positive peak at negative time lag was relatively larger, and the negative peak was negligibly smaller. This could be due to the fact that the HP location is not precisely at the peak pressure point (due to experimental constraints), but is rather at a location in the region of pressure rise from the pressure minimum (at the low-pressure region) to the peak pressure point. In the region of rising pressure, the correlation trends are highly sensitive to the location, since it is a region where a shift in the sense of correlation peak (from positive to negative) occurs, due to the stretching/compression of the centerline pressure profile as explained by \cite{brusniak1994physics}. Therefore, the observed differences in MS-HP correlation could either be due to small differences in the extent of the high-pressure zone between experiments and computations, or possibly because of the finite sensing area of the transducer in experiments as opposed to the computations in which the data is collected at a single grid point. The fluctuations at the peak pressure point (downstream of HP) in computations exhibited negative correlation with the fluctuations at MS, which was observed for the cuboid case having a larger extent of high-pressure region (refer to figure \ref{figure12:statistical relation-MS(spanwise)-HP-experiments-shortcylinder-cuboid}(b)). Nonetheless, the trends in correlations between MS and other regions on the wall at the centerline are consistent with the expected trends based on the pressure profile distortions due to shock motion.

The second discrepancy is with regard to the correlation between mean shock foot (MS) and reattachment ($RA$). In computations, a negative peak correlation (denoted as P4) of around 0.2 can be noted between MS and $RA_{C}$ in figure \ref{figure12:centerline-statistical relation-MS-PP-HP-RR}, whereas the MS - $RA_{E}$ correlations were shown to be negligible. Understandably, it is due to the experimental limitation in mounting the transducer at the precise reattachment point for the cubical protuberance. For the cuboid case, a large negative peak correlation of $\sim 0.35$ was observed in the experimental MS - $RA_{E}$ correlations (see figure \ref{figure9:correlation-sep-reat-experiments}), which is consistent with the observation between MS and precise reattachment point in the computations. The other noteworthy aspect in the MS - $RA_{C}$ correlation is the small peak at $St_{\delta}$ of $\sim 0.7-0.9$, suggesting the relation between the two locations through the vortices being shed.

\begin{figure*}[!htb]
    \centering
    \includegraphics[width=0.95\textwidth]{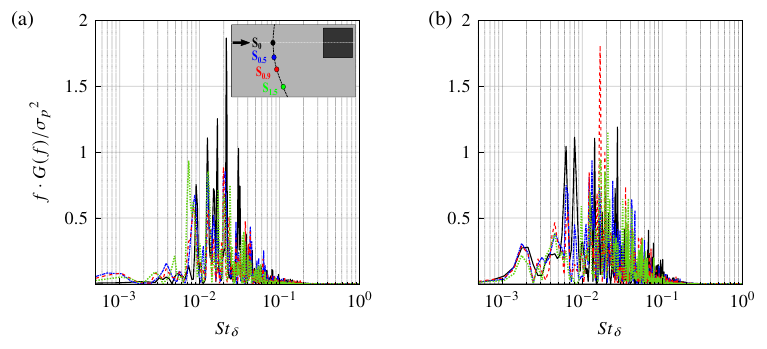}
    \caption{Power spectral densities of probes along the mean shock (a) R1\_SPSW (b) R2\_SPOF}
    \label{figure13:spanwise-probes-psd}    

\end{figure*}

\begin{figure*}[!htb]
    \centering
    \includegraphics[width=0.95\textwidth]{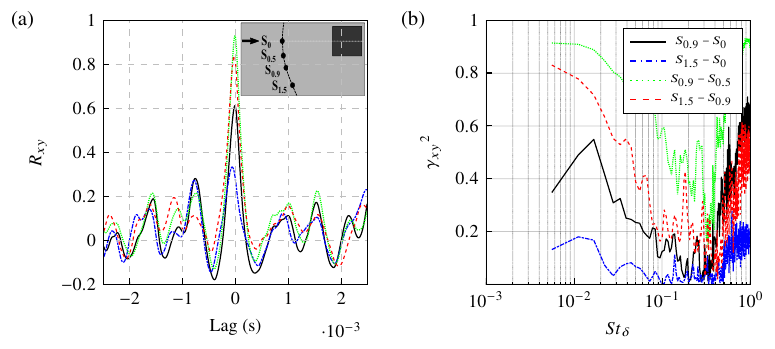}
    \caption{Statistical relations among probes along the mean shock foot (a) cross correlation. (b) coherence.}
    \label{figure14:statistical relation-spanwise-probes}     
\end{figure*}

The trends in spectra, correlations and coherence along the mean shock foot at different spanwise locations are again similar to the experimental observations (see figure \ref{figure10:correlations-mean-shock-shortcylinder-cuboid}). The PSD of signals at different spanwise locations along the mean shock foot is shown in figure \ref{figure13:spanwise-probes-psd} for both slip wall and outflow cases. At all locations, the low frequencies are found to be dominant. Significant two-point correlations are observed between all locations, as shown in figure \ref{figure14:statistical relation-spanwise-probes}(a), and the values are especially high between the pairs that are away from the spanwise centerline ($S_{1.5}-S_{0.9}$ and $S_{0.9}-S_{0.5}$). Negative time lags are observed in the correlations of pressure fluctuations between the centerline and other locations away from it, as was also observed in experiments, suggesting a flapping motion of the shock foot. Significant coherences are observed between all locations at low frequency ranges (figure \ref{figure14:statistical relation-spanwise-probes}(b)), suggesting that the shock foot exhibits a coherent to-and-fro motion at low frequencies ($St_{\delta} \sim 0.01$), along with a flap about the spanwise center. Trends in correlation between pressure fluctuations at HP and at different spanwise mean shock foot locations is also similar to those observed in experiments. The complex shock foot motion shall be discussed further in the following section.

%%%%%%%%%%%%%%%%%%%%% Statistical relation:: MEAN SHOCK && PLATEAU PRESSURE %%%%%%%%%%%%%%%%%%%

\FloatBarrier

\subsection{Visualization of spanwise shock motion}
\label{Sec4.5:spanwise-shock-motion}
\begin{figure*}[hbt!]
    \centering
    \includegraphics[width=0.80\textwidth]{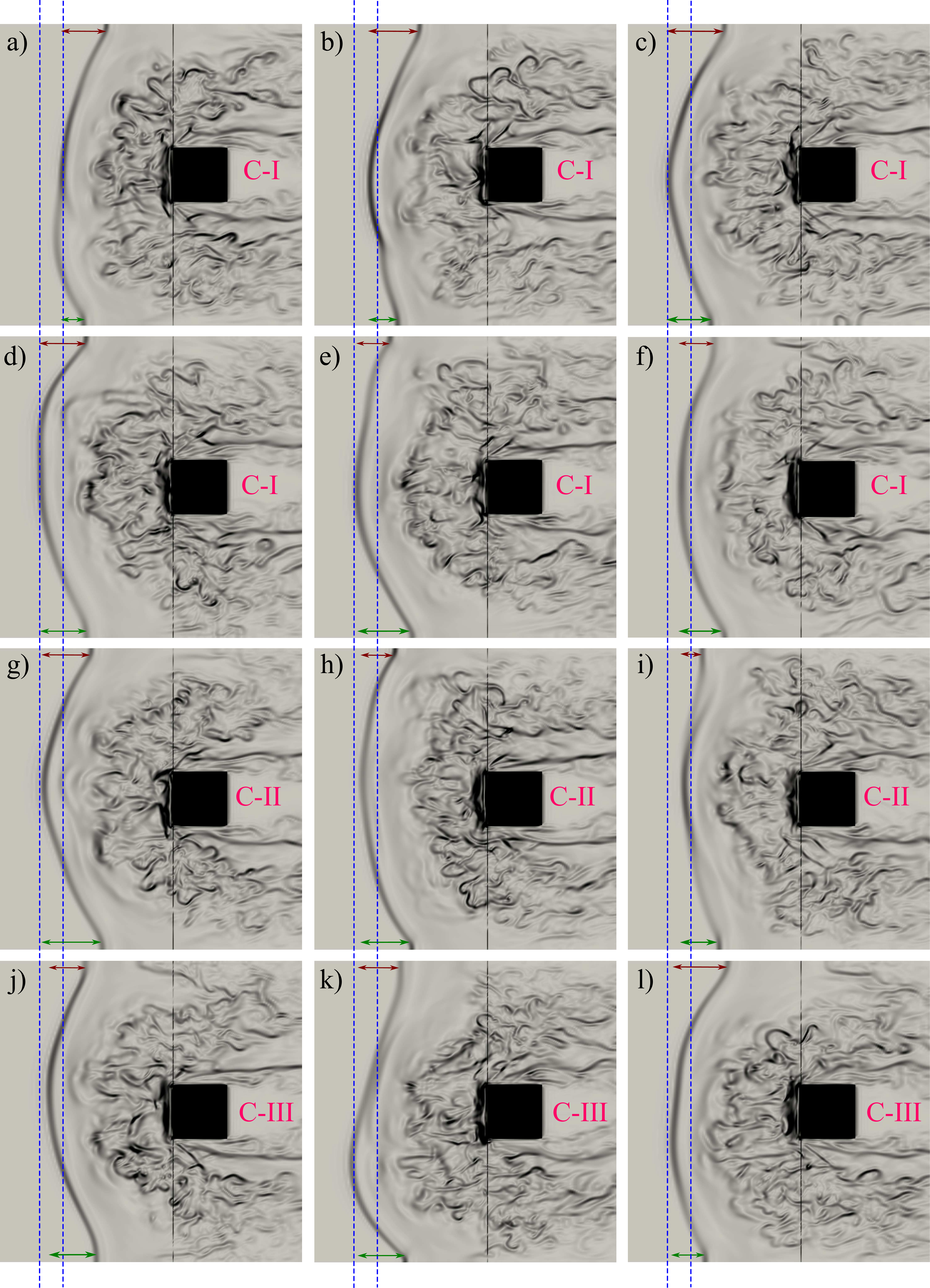}
    \caption{Instantaneous picture of spanwise shock motion at different instances. The dotted lines represent upstream-most and downstream-most positions corresponding to three cycles.}
    \label{figure16:temporal-evolution of spanwise-shock}
\end{figure*}

From the top view ($x-z$ plane), the separation shock foot was observed to exhibit complex movements which are asymmetric about the spanwise centerline in most instances, rather than a simple back and forth motion as a whole with its mean shape intact. Such a motion is illustrated in the figure \ref{figure16:temporal-evolution of spanwise-shock} through a sequence of instantaneous snapshots ($x-z$ plane) taken at $y/\delta = 0.5$. The snapshots constitute three consecutive cycles, with the first two rows of snapshots corresponding to the first cycle (C-I), while the subsequent rows correspond to the second (C-II) and third cycles (C-III), respectively. The two dotted lines represent the upstream-most and downstream-most positions corresponding to those three cycles. The streamwise distance between the separation shock position at the spanwise center and the shock positions at the spanwise ends is shown in red and green color lines, respectively. The instantaneous distance between the spanwise ends of the shock and any one of the two dotted lines may also be referred to follow the local shock foot motion. 

In the first snapshot, figure \ref{figure16:temporal-evolution of spanwise-shock}(a), the separation shock foot is at the downstream-most position. The shock foot is asymmetric about the spanwise center, with the shock foot position at the right end of the spanwise centerline (the upper end of the figure) significantly downstream of the shock foot positions at the left side of the centerline. The shock foot is relatively flat, especially to the left side of the centerline. In the next instance (figure \ref{figure16:temporal-evolution of spanwise-shock}(b)), the upstream motion of the shock is initiated in the vicinity of the centerline. The shock foot can be seen to reach the upstream-most position close to the centerline in the next instance, shown in figure \ref{figure16:temporal-evolution of spanwise-shock}(c). The shock foot exhibits a relatively larger curvature near the centerline, which is aided by the inertia of shock foot locations considerably away from the centerline, as apparent from the relatively fixed shock foot positions at the spanwise ends. In the subsequent instance (figure \ref{figure16:temporal-evolution of spanwise-shock}(d)), the shock foot tends to flatten, especially in the right half of the span aided by the upstream movement of the shock foot in the right half. At this instance too, the shock foot is asymmetric, but with the right half relatively upstream of the left half, in contrast to the sense of asymmetry in the very first snapshot. The next instance (figure \ref{figure16:temporal-evolution of spanwise-shock}(e)) marks the initiation of the downstream retreat of the shock foot. Although the shock foot near the centerline starts to move downstream, it is worth noting that the shock foot at the right end continues its upstream movement from the previous instances, whereas the shock foot at the left end moves downstream. In the subsequent instance (figure \ref{figure16:temporal-evolution of spanwise-shock}(f)), which marks the end of the first cycle, the centerline shock foot can be seen to have reached the downstream-most position. The shock foot is again flatter, and is relatively symmetric at this instance. In the next cycle C-II (figure \ref{figure16:temporal-evolution of spanwise-shock}(g)-(i)), during both upstream and downstream movements, the shock foot is apparently symmetric about the centerline for most instances in this cycle, but towards the end of the cycle, when the shock foot reaches the downstream-most position, it exhibits asymmetry when the next cycle C-III is initiated. The cycle C-III (figure \ref{figure16:temporal-evolution of spanwise-shock}(j)-(l)) exhibits asymmetrical motion of the shock foot similar to C-I, but the asymmetries in this cycle are in the opposite sense to that in C-I. 

%%%%%%%%%%%% Spanwise shock features %%%%%%%%%%%%%

\begin{figure*}[!htb]
    \centering
    \includegraphics[width=0.95\textwidth]{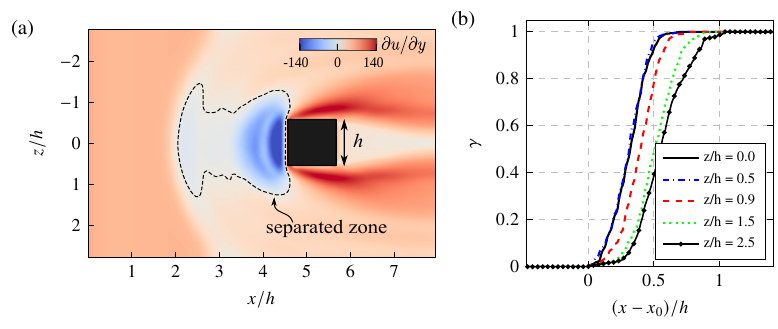}
    \caption{Distribution of (a) mean streamwise velocity gradient at the wall (dotted curve indicates the separated zone). (b) Intermittency ($\gamma$) of separation shock}
    \label{figure17: mean-sep-zone-intermittency}
        
\end{figure*}

The following are the common features in all the cycles: 1. The upstream or downstream motion of the shock foot is initiated in the neighborhood of spanwise centerline, followed by shock foot at locations away from centerline, and 2. Due to this lag, the upstream movements are marked by larger curvatures of shock near the centerline, and the downstream movements are marked by flattening of the shock foot. Thus, in addition to the to-and-fro motion, the shock foot also exhibits a flapping motion about the centerline, which is often asymmetric, even anti-symmetric, considering that the sense of asymmetry gets swapped between the cycles. Due to this swap, as well as the presence of some cycles exhibiting relatively symmetrical shock motion, the frequencies associated with the anti-symmetry can be expected to be lower (roughly one third) of the frequencies associated with the to-and-fro motion. This will be evident when we present the spectra associated with POD modes of wall pressure fluctuations. The observation that the upstream/downstream motion of the shock is initiated in the neighborhood of the centerline can be related to the spanwise extent of the reverse flow above the wall in the separated region, since the motion is associated with the reverse flow mass excess/deficit. Figure \ref{figure17: mean-sep-zone-intermittency}(a) shows the mean reverse flow region ($du/dy < 0$), marked by a dotted curve, extending approximately $1.5 h$ on either side of the centerline. Therefore, unsteady flow fields in this zone directly influence the dynamics of the separation shock. The asymmetries are caused by the instantaneous skewness in the reverse flow direction about the mean. The intermittency of the separation shock position at different spanwise locations is also shown in figure \ref{figure17: mean-sep-zone-intermittency}(b). While the intermittent length scales are not very different near the centerline for a spanwise extent of about $\pm 0.5h$, it increases with spanwise distance as we go further away from the centerline, which is indicative of the flapping motion of the shock foot about the centerline. For the simulations with outflow boundary conditions at the spanwise ends (R2\_SPOF), the observations with regard to shock motion are similar, but the motion was often symmetric, and the anti-symmetries were relatively less dominant. It is interesting to note that the skewness and anti-symmetry are more pronounced in the presence of wall confinement. These observations shall be examined further by means of modal analysis.

\subsection{Proper orthogonal decomposition}
\label{Sec4.6:POD}
%% mid-span schlieren pod
Modal analysis of the flow field is conducted through proper orthogonal decomposition (POD) to identify the energetically dominant structures (modes) and corresponding frequencies. Numerical density gradients (schlieren) along the mid-span plane, and pressure fields on the bottom wall are used to explore the most energetic activities occurring in the in the respective plane. Additionally, POD based on three-dimensional velocity data is also performed to get more insights about the dominant phenomena.

\FloatBarrier  % floats in this subsection end here

%%% Mean Surface Pressure Distribution:: Experiment and Computation at z = 0 (centerline) %%%

%%%%%%%%%%%%%%%%%%%%%%%%%%%%%%%%%%%%%%%%%%%%%%%%%%%%%%%%%%%%%%%%%%%%%%%%%%%%%%%%%%%%%%%%%%%%%%%%%%%%%%%%%%%%%%%%%

POD features derived from numerical schlieren data in the mid-span plane are qualitatively similar to those observed in experiments. The first two modes, both in experiments and in computations, exhibit low-frequency separation shock oscillations as the dominant phenomenon, whereas higher modes are characterized by smaller structures at mid- and higher frequency ranges. Therefore, the mid-span POD from the computations does not offer new insights than those from the experimental results, and only serves as a further validation. However, with the computational data, a comprehensive understanding of interaction dynamics can be obtained by looking at the dominant phenomena at other planes, which cannot be accessed from experiments. 
 
\subsubsection{Two-dimensional POD: wall pressure}
\label{Sec4.6.2:wallpressure-POD}
\begin{figure*}[!htb]
    \centering
    \includegraphics[width=0.95\textwidth]{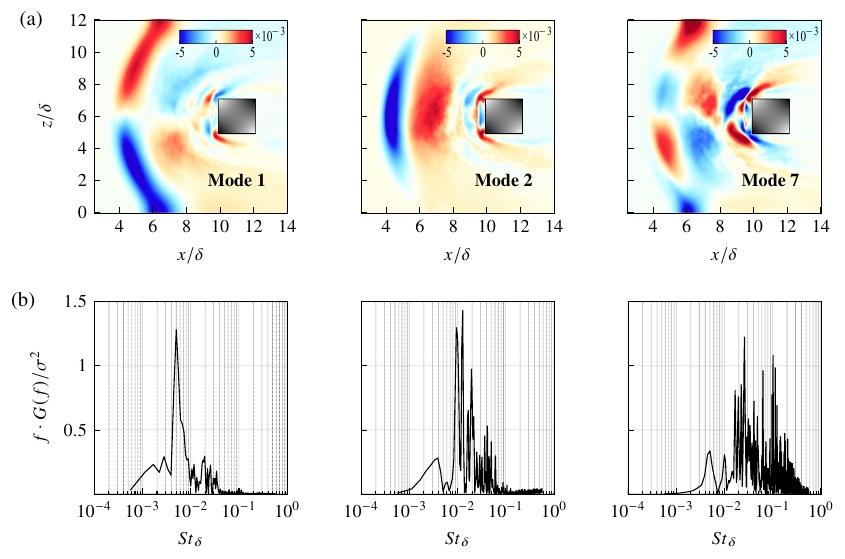}
    \caption{POD based on wall pressure corresponding to mode 1, 2 and 7 (R1\_SPSW). (a) spatial mode. (b) PSD of temporal coefficients of the respective mode}
    \label{figure19:wall-pressure-pod-initial-modes}
 \end{figure*}

POD based on wall pressure data helps in understanding the shock foot motion along the span and its statistical relations with pressure fluctuations in other zones in the interaction. The dominant structures and corresponding frequencies associated with spatial modes are shown in figures \ref{figure19:wall-pressure-pod-initial-modes}(a) and (b). The first mode presents a nearly antisymmetric (rather than asymmetric) shock foot motion, while the coherent and symmetrical to-and-fro shock oscillations are apparent only in the second mode. Both the modes have comparable energies and exhibit dominant amplitudes in low-frequency bands. It is noteworthy that the dominant frequency in the first mode, corresponding to anti-symmetric oscillations, is $St_{\delta} \sim 0.005$, which is considerably lower than the frequencies ($St_{\delta} \sim 0.01-0.02$) associated with the second mode, corresponding to the to-and-fro motion of shock foot. This is consistent with the remarks concerning the different phases made earlier in reference to Figure \ref{figure16:temporal-evolution of spanwise-shock}.
The mean shock foot sweep in both these modes exhibit a considerable relation with the wall pressure oscillations in the vicinity of protuberance base. In the second mode, it is observed that the separation shock exhibits an alternating sense of correlation with other regions of the interaction, a trend which was discussed earlier in connection with the correlations in pressure fluctuations along the centerline. To begin with, a strong negative correlation is observed between the shock and plateau pressure regions, and the correlation switches to positive values, albeit weak in the region of pressure rise downstream of the pressure minima locations, and again turns negative in the high-pressure region. The seventh mode presents organized large scale spanwise distortions in shock foot and related structures in the separated region. The mid-frequencies start becoming prominent from this mode, suggesting that mid-frequency phenomena could partly be associated with large-scale spanwise oscillations, apart from shear layer shedding. Higher modes are characterized with frequencies ranging from mid-level ($St_{\delta} \sim 0.1$) to high-level ($St_{\delta} \sim 1$) as the associated scales of the activities become smaller and smaller. 

\begin{figure*}[!htb]
    \centering
    \includegraphics[width=0.95\textwidth]{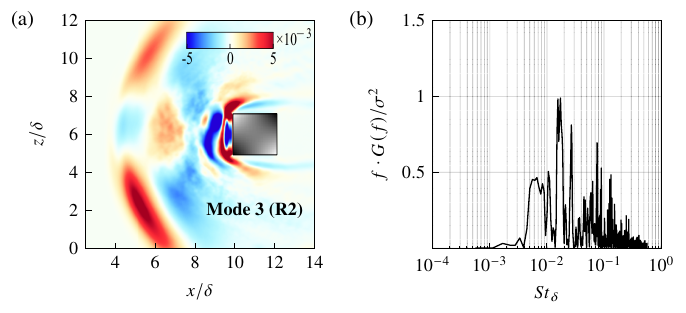}
    \caption{POD based on wall pressure (R2\_SPOF). (a) spatial mode. (b) PSD of temporal coefficients}
    \label{figure20:wall-pressure-pod-higher-modes}
    
 \end{figure*}

For the R2\_SPOF simulations too (with outflow conditions at spanwise boundaries), qualitatively similar features are generally observed in the POD of surface pressure fluctuations, except that the first two modes are swapped. In the R2\_SPOF case, the first mode corresponds to the to-and-fro symmetric shock motion while the second mode corresponds to the anti-symmetric flap with relatively lower frequencies than the to-and-fro oscillations. It is interesting to note that the relieving effects at spanwise ends with the outflow boundary condition seem to weaken the anti-symmetric movements. A related noteworthy aspect in the POD of the R2\_SPOF case was that third mode presents a symmetric flapping of the shock foot about the centerline, as shown in figure \ref{figure20:wall-pressure-pod-higher-modes}, with a considerable correlation between the fluctuations in the high-pressure zone and the shock foot away from centerline. Such large scale symmetrical flaps were not apparent in any dominant POD mode with slip wall conditions. 

It may be recalled from section \ref{sec2.4:corr-coher-experiments} (with reference to figure \ref{figure12:statistical relation-MS(spanwise)-HP-experiments-shortcylinder-cuboid}), that the correlations with the fluctuations at high-pressure zone near protuberance base were significant, not just for fluctuations at the mean shock foot location at centerline, but also for other spanwise locations along mean shock foot. Particularly, for the case of cubical protuberance at similar flow conditions, \cite{ramachandra2023} reported that the correlations with pressure fluctuations near protuberance base for the mean shock locations at other spanwise locations (as far as $1.5 h$ from centerline) had peak values which were even higher than the peak value for centerline mean shock foot location. The observations from the POD corroborate the trends observed in correlations in surface pressure between high-pressure zone and the mean shock foot, suggesting that the fluctuations near the protuberance base is related to both the to-and-fro motion and the flapping motion (symmetric as well as anti-symmetric) of the shock foot.

%%%%%%%%%%%%%%%%%%%%%%%%%%%%%%%%%%%%%%%%%%%%%%%%%%%%%%%%%%%%%%%%

%%%%%%%%%%%%%%%%%%%%%% MODE-1, 2 and 3 %%%%%%%%%%%%%%%%%%%%%%%
 %%%%%%%%%%%%%%%%%%%%%%%%%%%%%%%%%%%%%%%%%%%%%%%%%%%%%%%%%%%%%%%%%%%%%%%%%%

%%%%%%%%%%%%%%%%%%%%%% MODE-10, 100 and 3(R2) %%%%%%%%%%%%%%%%%%%%%%%
 %%%%%%%%%%%%%%%%%%%%%%%%%%%%%%%%%%%%%%%%%%%%%%%%%%%%%%%%%%%%%%%%%%%%%%%%%%
 
 %\FloatBarrier
 %%%%%%%%%%%%%%%%%%%%%%%%%%%%%%%%%%%%%%%%%%%%%%%%%%%%%%%%%%%%%%%%%%%%%%%%%%%%%%%%%

\subsubsection{Three-dimensional POD: velocity field}
\label{Sec4.6.3:three-dimensional-POD}
\begin{figure*}[!htb]
    \centering
    \includegraphics[width=0.90\textwidth]{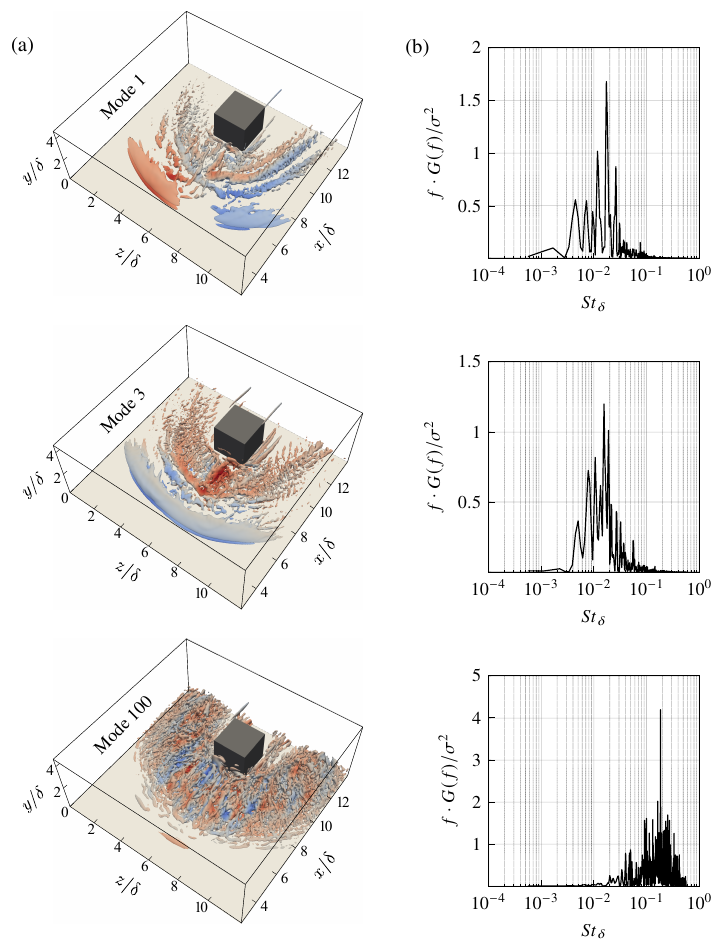}
    \caption{Three-dimensional POD: (a) spatial mode (b) PSD}
    \label{figure21:three-dimensional-pod-1-2-5-modes}
 \end{figure*}

The computational data also facilitate a detailed analysis of the three-dimensional picture. A three-dimensional POD of the velocity data is conducted to resolve the dominant three-dimensional flow structures that are associated with the low-frequency shock motion. Three-component (u, v, w) velocity data is used to obtain the modes and corresponding frequencies. Few selected modes and their spectra for the R2\_SPOF case are presented in figure \ref{figure21:three-dimensional-pod-1-2-5-modes}. As discussed earlier, the interaction features from both simulations (R1\_SPSW and R2\_SPOF) are qualitatively similar and agree well in several aspects. Nonetheless, simulations with spanwise-slip wall conditions (R1\_SPSW) resulted in more complex interaction features due to spanwise confinement, leading to skewing of the horseshoe vortex towards one side, along with pronounced anti-symmetric effects. Therefore, to understand the essential features of the interaction and the underlying mechanism responsible for shock unsteadiness, the 3-D POD and reduced-order reconstruction with dominant modes based on simulations with spanwise-outflow (R2\_SPOF) conditions is considered. In the 3-D POD with the velocity data, the first six modes represent separation shock and horseshoe vortex as dominant events exhibiting low frequencies, i.e., $St_{\delta} \sim 0.01$; the first and second modes capture anti-symmetric shock oscillation, while the third mode presents the to-and-fro motion. It is observed that beyond mode six, mid-frequency ($St_{\delta} \sim 0.1$) events start to appear and become dominant at higher modes, for instance, mode 100 showing a clear peak centered at $St_{\delta} \sim 0.1$. Further higher modes are related to finer scales of turbulence-like structures at $St_{\delta} \sim 1$.

%%%%%%%%%%%%%%%%%%%%%%%%%%%%%%%%%%%%%%%% 3-DIMENSIONAL POD MODES && PSD %%%%%%%%%%%%%%%%%%%%%%%%%%%%%%%%%%%%%%%%%

%%%%%%%%%%%%%%%%%%%%%% MODE-1, 2 and 3 %%%%%%%%%%%%%%%%%%%%%%%

 %%%%%%%%%%%%%%%%%%%%%%%%%%%%%%%%%%%%%%%%%%%%%%%%%%%%%%%%%%%%%%%%%%%%%%%%%%

 %\FloatBarrier  % floats in this subsection end here
 %%%%%%%%%%%%%%%%%%%%%%%%%%%%%%%%%%%%%%%%%%%%%%%%%%%%%%%%%%%%%%%%%%%%%%%%%%%%%%%

\subsection{Flow reconstruction with low-frequency modes}
\label{sec4.7:flow-reconstruction}
\begin{figure}[hbt!]
    \centering
    \includegraphics[width=0.85\textwidth]{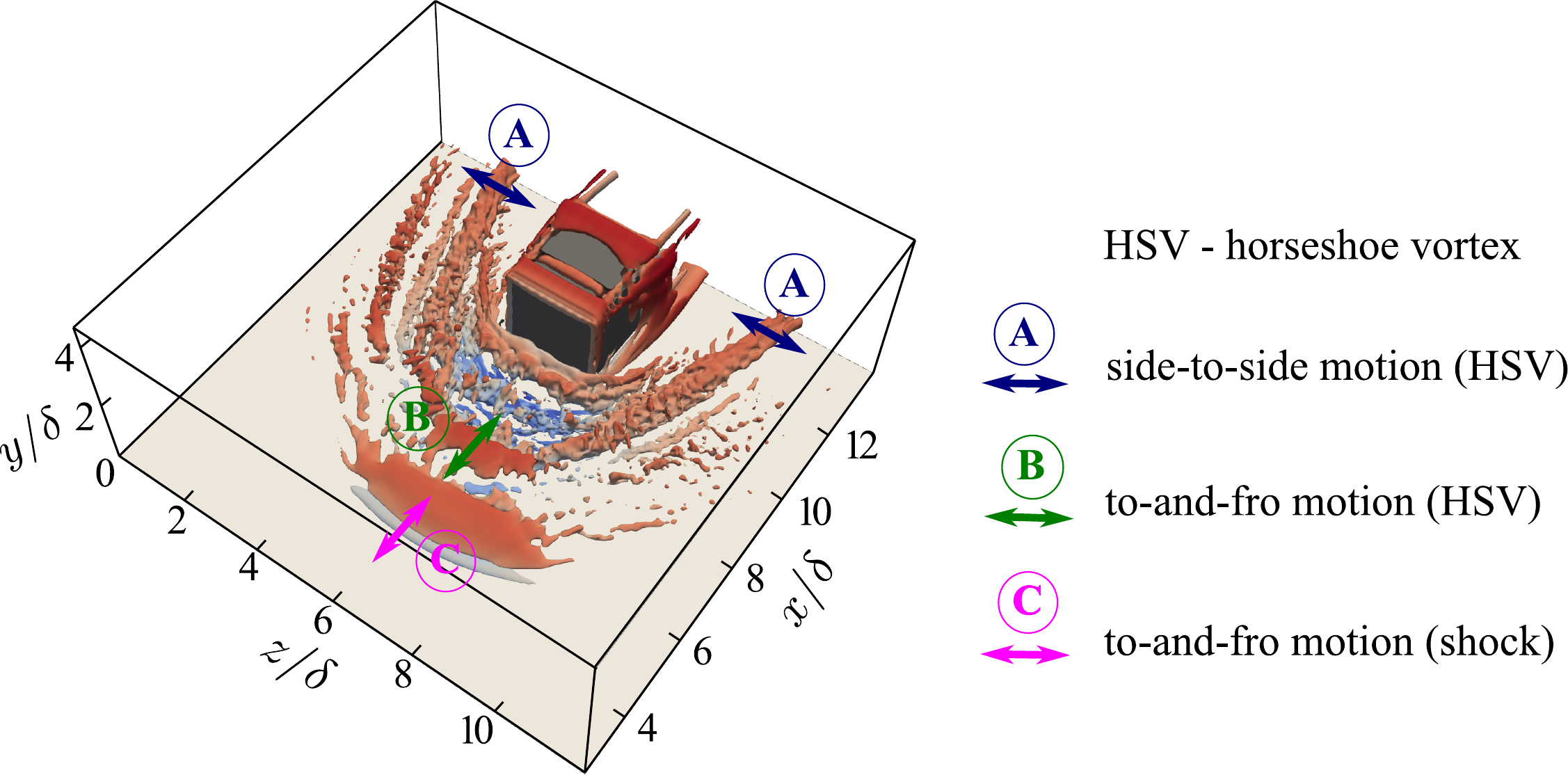}
    \caption{Reconstructed flow field with low-frequency modes.}
    \label{figure23:reconstructed-flow}
 \end{figure}

Owing to the complexities of interaction, a reduced-order modeling and reconstruction of the flow field is performed using the first six 3-D POD modes, which constitute low-frequency events, to narrow down the focus on the source of shock motion. An instantaneous snapshot of the reconstructed flow field is shown in the figure \ref{figure23:reconstructed-flow}; the directions of motion of different features, such as shock and horseshoe vortex are also marked. The temporal evolution of the reconstructed flow field reveals that the central portion of the horseshoe vortex undergoes to-and-fro motions, influencing the movements of the separation shock. The horseshoe vortex exhibits spanwise wagging (sideway) motions around the protrusion, which is essentially inferred from the intercept of the ends of the vortex at the (downstream) outlet plane. It is observed that the spanwise movements of the horseshoe vortex are linked to the activities of the horseshoe vortex at the central portion, especially the radius of curvature (ROC) of the horseshoe vortex at the spanwise centerline upstream of the protuberance. To illustrate this, a sequence of instantaneous reconstructed views of the flow during a cycle of oscillation is shown in figure \ref{figure24:Evolution of reconstructed flow}. The left side of these figures shows the Q-criterion iso surfaces, while the right side presents the streamwise velocity field in the wall-parallel plane at $y/{\delta} = 0.4$ to elucidate the explanation.

To begin with, an instance (t = 0.45 ms) where the radius of curvature of the horseshoe vortex at the central portion is large is provided. At this instance, the separation shock is at its downstream-most position, and the horseshoe vortex ends are far apart from each other. In the next instance, the radius of curvature of the central portion decreases, as the separation shock is pushed upstream and the ends of the horseshoe vortex at the outlet plane move towards each other. In the subsequent instances, the radius of curvature of the central horseshoe vortex decreases further and becomes minimum at $t = 0.91$ ms (see figure \ref{figure24:Evolution of reconstructed flow}). During this time period, the separation shock moves forward and assumes its upstream-most position, whereas the horseshoe vortex ends came closer and closer, eventually reaching a position where they are closest to each other. This completes half a cycle. It may be noted that while the shock foot moves upstream in the neighborhood of the centerline, at spanwise locations away from the centerline, the shock foot does not move upstream considerably, distorting the shape of the shock. The instances after $t=0.91$ ms show the retraction/downstream moving phase of the separation shock. At $t = 1.06$ ms, when the centerline shock foot is seen to start moving downstream, it may also be noted that the shock is seen to have moved upstream near the spanwise ends. As the shock moves downstream, the radius of curvature of the horseshoe vortex at the spanwise center is seen to increase, with the front portion (upstream of protuberance) of the horseshoe vortex becoming flatter, and the downstream ends of the horseshoe vortex in the outlet plane moving away from each other. The last instance shows the scenario where the radius of curvature becomes large (similar to the first instance). At this instance, the separation shock reaches its downstream-most position, and the horseshoe vortex ends are at the farthest position from each other, before the next cycle begins.

\begin{figure*}[!htb]
    \centering
    \includegraphics[width=0.90\textwidth]{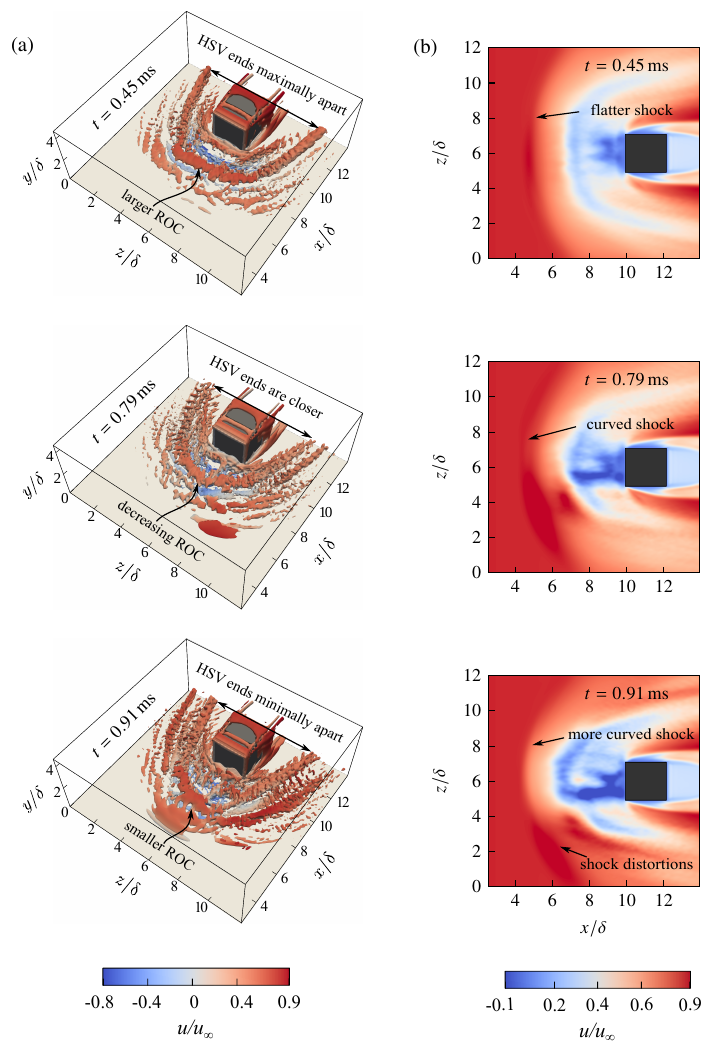}
    \caption{Evolution of reconstructed flow field. (a) Q-criterion. (b) streamwise velocity at $y/\delta = 0.4$. }
    \label{figure24:Evolution of reconstructed flow}
 \end{figure*}

%%%%%%%%%%%%%%%%%%%%%%%%%%%%%%%%%%%%

 \begin{figure*}[!htb]
    \ContinuedFloat         %% to make the current plot a continuation of the previous one.
    \centering
    \includegraphics[width=0.90\textwidth]{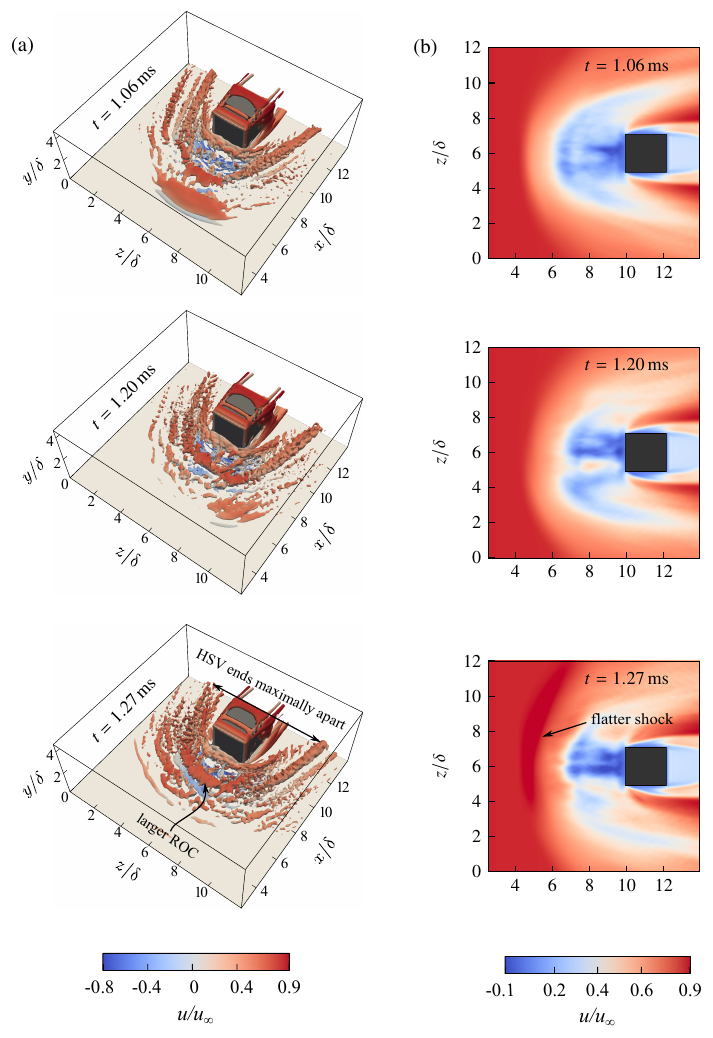}
    \caption{Evolution of reconstructed flow field (continued). (a) Q-criterion. (b) streamwise velocity at $y/\delta = 0.4$.}
    \label{figure25:Evolution of reconstructed flow}
 \end{figure*}

% \FloatBarrier  % floats in this subsection end here
 %%%%%%%%%%%%%%%%%%%%%%%%%%%%%%%%%%%%%%%%%%%%%%%%%%%%%%%%%%%%%%%%%%%%%%%%%%%%%%%%%%%%%%%%%%%%%%%%%%%%%%%%%%%%%%%%

\subsection{Discussion: Mechanism of low-frequency dynamics}
\label{sec4.8:governing-mechanism}
From the discussions in the previous section, two important and related observations give some
valuable insights on the mechanism sustaining the low-frequency unsteadiness. First, the shock foot
motion (both upstream and downstream) is initiated in the vicinity of the spanwise centerline, and the
motion of the shock foot away from the centerline initiates after the centerline shock foot gets displaced.
Second, associated with shock motion, the horseshoe vortex exhibits a complex dynamics, involving
to-and-fro motion and distortions. Particularly, the radius of curvature of the horseshoe vortex upstream
of the protuberance changes, decreasing or increasing as the centerline shock foot moves upstream or
downstream, respectively. Along with that, the trailing ends of the horseshoe vortex (in the outlet
plane) move closer to or away from each other, respectively with decrease or increase in the centerline
radius of curvature. These low-frequency oscillations relate to other important observations in
experiments and computations, which were discussed in the earlier sections.

The pressure oscillations near the protuberance base (which are below the reattachment point on the protrusion face) exhibit good correlation with the shock foot motion all along the span, even influencing the symmetric and anti-symmetric motion of the shock about the spanwise centerline. The reverse flow emanates from the high-pressure zone near the protuberance base, and it directly impacts the shock foot for a certain spanwise extent in the neighborhood of the centerline, depending on the spanwise extent of the reverse flow along the separation line. When the shock foot is in its downstream-most position during a cycle, i.e., when the instantaneous (centerline) separation length is the shortest for the cycle, the reattachment pressure as well as the peak pressure near the protuberance base are high, assuming the maximum value for the particular cycle. This results in an excess mass getting injected into the separated region from the reattachment location. The mass flow rate at different time instances shown in figure \ref{figure24:Evolution of reconstructed flow} are estimated in a plane parallel to the wall at a height of $0.5\delta$ above the wall, the span of the plane aligning with the span of the protuberance, with the plane upstream from protuberance base for a length of $0.8\delta$ (just downstream of the core of horseshoe vortex in the mean flow). At $t = 0.45$ ms, when the shock foot is in its downstream-most position, the downward mass flow rate on the plane is 0.0092 $kg/s$. As the shock foot moves upstream, the pressure profile stretches and the pressure at the reattachment as well as near the protuberance base drops based on correlation trends (which was evident from the trends in the correlations), and the mass flow rate through the plane is observed to reduce. When the shock is at its upstream-most position ($t = 0.91$ ms - $1.06$ ms), mass flow rate is observed to be the minimum for the cycle, which is 0.0077 $kg/s$. As the shock foot at the centerline moves downstream, the mass flow rate through the plane is also observed to increase, reaching a value of 0.0093 $kg/s$ when the shock foot reaches the downstream-most position again before the next cycle. This excess or deficit mass in the reverse flow is directly related to the shock foot motion close to the centerline. 

\begin{figure*}[!htb]
    \centering     
    \includegraphics[width=0.70\textwidth]{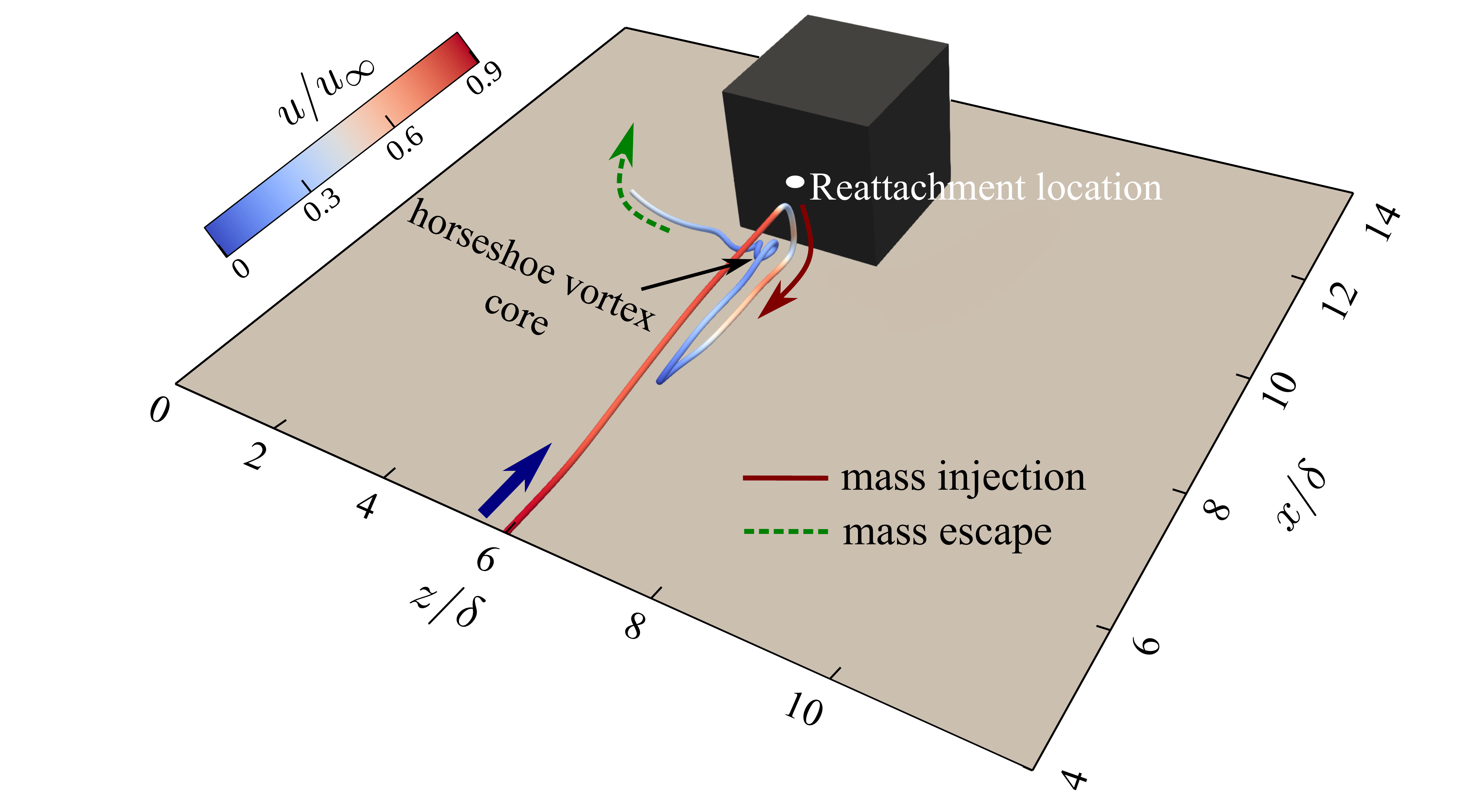}
    \caption{A streamline in the mean flow field representing the mean path of a particle moving along the separated shear layer.}
    \label{figure26:streamline-time-calculation}
\end{figure*}

However, in the 2-D separation which consists of a closed separation bubble, entrainment is the only means other than injection at reattachment by which the bubble transacts mass with the outer flow. In the present case of open separation, the mass that is injected at the reattachment goes around the horseshoe vortex, and eventually escapes out spanwise through the horseshoe vortex. This is illustrated in figure \ref{figure26:streamline-time-calculation} by means of a streamline along the spanwise center in the mean flow field. Thus, any excess mass that is injected will eventually relieve out of the separated region through the horseshoe vortex, and for the excess mass to reach the horseshoe vortex core it takes a certain time delay after it is injected at the reattachment. The flow speeds along the streamline are shown using a color map in the figure. The time taken by a particle moving along the streamline is determined from the total distance travelled along the streamline and the corresponding velocity (averaged over the streamline). The average velocity is calculated by integrating the velocity values along the path and then dividing by the total length. From these, the time scale associated with the particle motion along the streamline is estimated to be around 0.312 ms, which corresponds to $St_{\delta} \sim 0.0365$, which is in the same orders of the low-frequency shock motion. 

Further, one could split the path into roughly two equal parts- one, from reattachment to the separation line; and two, from the shock foot to the horseshoe vortex core. The excess mass after reaching the separation line turns back towards the horseshoe vortex core. As the excess mass reaches the core, the spanwise outward mass flow is large, resulting in stretching of the horseshoe vortex at the centerline, which in turn increases the radius or curvature of the horseshoe vortex, pushing the trailing ends of the horseshoe vortex away. It may be recalled from figure \ref{figure24:Evolution of reconstructed flow} that when the centerline separation length is shortest, the radius of curvature of the horseshoe vortex at the centerline is also the largest for the cycle. At the same time, when there is an excess mass injection at reattachment, the excess mass from the previous cycle gets drained through the horseshoe vortex core. On the other hand, when the centerline separation is in its upstream-most position, the excess mass is turned back upstream; the excess mass from the previous cycle has drained, and the mass flow through the horseshoe vortex core is lesser, which in turn results in a smaller radius of curvature due to reduced spanwise stretching. Since the distance that the excess mass must travel from reattachment to the horseshoe vortex core is of the orders of the separation length, the centerline separation length is the appropriate scale for the low-frequency dynamics, as observed from the scaling for spectra of shock foot pressure oscillations from experiments with various protuberances.  

In summary, it is inferred that the mechanism sustaining the low-frequency shock oscillations for the
three-dimensional SBLI due to protuberances is essentially an imbalance and time lag between the
mass injection at the reattachment on the protuberance face and the mass escaping (relieving)
spanwise through the horseshoe vortex. The spanwise extent of reverse flow, which in turn is related
to the centerline separation length, determines the amplitude and spanwise coherence of the to-and-fro
motion. Since it is the spanwise extent of the reverse flow that the to-and-fro motion is driven, a
flap of the shock foot about the centerline is generated, which results in larger amplitudes of shock
foot oscillations away from the centerline.

\section*{Acknowledgements}
The authors would like to acknowledge the National Supercomputing Mission (NSM) for providing access to computing resources of PARAM RUDRA at P G Senapathy Center, Play Field Avenue, Indian Institute of Technology Madras (IIT Madras), Tamil Nadu 600036, which is implemented by C-DAC and supported by the Ministry of Electronics and Information Technology (MeitY) and Department of Science and Technology (DST), Government of India. The authors are also grateful to G. Rajesh and his group at the Department of Aerospace Engineering, IIT Madras, for helping with some equipment for experiments. The experimental work was supported by the Science and Engineering Research Board (SERB) of the Department of Science and Technology, Government of India, SERB Grant No. SRG/2019/001793.

\section*{Declaration of interests}
The authors report no conflict of interest

 \FloatBarrier  % floats in this subsection end here
 %%%%%%%%%%%%%%%%%%%%%%%%%%%%%%%%%%%%%%%%%%%%%%%%%%%%%%%%%%%%%%%%%%%%%%%%%%%%%%%%%%%%%%%%%%%%%%%%%%%%%%%%%%%%%%%%

\bibliographystyle{jfm}
\bibliography{jfm}

\end{document}